**Disordered Continuity: Programming Resolution-Independent Stochastic Metamaterials with Differentiable Anisotropic Property Distribution**

*Canhui Lin [a,†], Ke Xu [a,†], Chenli Zhou [a], Yubin Gao [a], Yingguang Li [a,*]*

a. Canhui Lin, Ke Xu, Chenli Zhou, Yubin Gao, Yingguang Li

College of Mechanical and Electrical Engineering, Nanjing University of Aeronautics and Astronautics, Nanjing, China

E-mail: liyingguang@nuaa.edu.cn

† Canhui Lin and Ke Xu contributed equally to this work.

Funding: This research was funded by the National Natural Science Foundation of China (Ref.: 52175466), and the Outstanding Youth Foundation of Jiangsu Province (Ref: BK20240184).

---

[*] Corresponding author. Email address: liyingguang@nuaa.edu.cn

## Abstract

Nature-inspired stochastic metamaterials with disordered and multiscale architectures have shown great promise towards extraordinary functionalities, including high mechanical resilience, stress modulation and biased stiffness reinforcement. As a special type of functionally graded metamaterial, programming multiscale stochastic metamaterial to achieve required functional property is computationally demanding due to the iterative simulation process, and thereby often intuitively implemented by filling a predefined subset of functional units into rasterized design space of fixed resolution, which restricts the flexibility and effectiveness of the designed functionality. To mitigate the computational complexity introduced by the multiscale architecture, we proposed a two-stage approach to programming stochastic metamaterials towards customized mechanical response. Instead of directly optimizing stochastic microstructures, the proposed approach first optimizes a differentiable physical property distribution, e.g. stiffness that targets desired functionality, and then generates spinodal architected microstructures to realize such property distribution under resolution-independent rasterization. The key enabler is the incorporation of spherical harmonics to represent, modulate and interpolate anisotropic stiffness distribution, which then serves as a non-uniform distribution function for the generation of anisotropic spinodal infills with high continuity. The test results demonstrated effective design of stochastic metamaterials with programmed functionalities to enable stress modulation, texture encoding and mechanical cloaking.

## 1. Introduction

Natural materials usually exhibit desired functionalities through millions of years' evolution. At microscopic level, natural materials are featured with disordered multiscale architectures (e.g. shells and wood) to enable unparalleled functionalities such as high resilience to external load [1-2], energy absorption [3-5], and spatially oriented deformation [6-7]. To mimic and even exceed beyond nature, stochastic metamaterials have drawn widespread attention for achieving graded and tailorable functional properties, which are potentially useful in aeronautical and biomedical applications [8-10]. Unlike periodic cellular metamaterials, e.g. chiral origami [11] whose unique functional properties are primarily determined by the constituent unit cells, stochastic metamaterials are architected by numerous anisotropic while continuously connected microstructures in varying degree of disorder. Therefore, the mechanical behavior (e.g. stress and displacement field) against external stimuli to realize certain functional property is predominantly programmed by the topology and spatial arrangement of the constituent microstructures [11-13].

The design of stochastic metamaterials for achieving the desired functional properties is typically a reverse design process which requires an iterative simulation process to establish a connection between the geometry and its corresponding mechanical behavior. However, simulation of multiscale geometry has long been acknowledged as a computation-demanding task due to a vast number of finite elements required for accurately depicting the intricate multiscale structures [14-15]. For an individual unit cell, the adoption of advanced machine learning algorithms as a surrogate model [16-20] can expedite the simulation process with high accuracy. However, learning the mechanical behavior of the entire stochastic structure is intractable for multiscale topology due to extremely high simulation cost. An alternative approach to optimizing the stochastic structure could be implemented by rasterizing the design space into finite number of voxels, each filled with a particular type of unit cell by *ad hoc* rules or machine learning schemes [21-23]. In this research direction, significant progress was made recently to facilitate the construction of stochastic metamaterials for various functional properties, such as collaborative design via topology optimization [24-26], implicit neural representations [27-28], and interpretable inversion models based on physical constraints [29-30], with respect to specific functional properties, e.g. stress modulation and mechanical cloaking.

Despite the promising progress achieved, the majority of recent optimization methods for stochastic metamaterials still adhered to a 'Lego-like' design principle, which required a rasterized design space with a predetermined and fixed resolution. Although adjacent unit cells can be seamlessly interconnected through special algorithmic treatment to eventually construct an integrated functional structure [31-32], the design flexibility and the target functionality are limited by the fixed resolution. Essentially, the mechanical behavior is uniquely determined by the distribution of effective physical property [29], e.g., the stiffness. Bearing this in mind, if a smooth stiffness tensor field can be constructed in advance to designing the geometry, it will help significantly reduce the complexity of the design problem. The underlying challenge is that stiffness and other physical properties, e.g. thermal conductivity, can be directionally variable and thus formulated as high rank tensor [33-34]. Optimizing tensor field throughout the design space using, e.g. gradient descend approach, would constantly require spatial interpolation for iterative update [35]. However, directly interpolating stiffness tensor gives rise to a distortional or biased distribution that undermines the optimization process. This poses an essential question regarding precise characterization and optimization to achieve differentiable stiffness distribution, as the key to programming resolution-independent stochastic metamaterials.

In this research, we incorporated spherical harmonics (SH) [36-37] to identify and optimize the stiffness distribution in a differentiable manner. As shown in **Figure 1**, the local stiffness tensors at critical positions of the design space are first determined for target functionality. These predefined stiffness tensors are represented by the coefficients of spherical harmonic basis functions, which theoretically guarantee the independent and precise modulation of stiffness tensors over the entire design space. Initially, a set of the original stiffness tensors represented as a 6*6 symmetric matrix are embedded in a highly nonlinear manifold, which are intractable for spatial interpolation. The transformation from the stiffness matrix into a decoupled SH coefficient matrix essentially converts the embedded manifold back into a Euclidean space, thus making the SH coefficient matrix differentiable and can be directly applied for smooth interpolation. Unlike previous metamaterial design methods that primarily focused on resolving the complication of individual unit cells, the incorporation of spherical harmonics proposed in this research pioneered a new direction for metamaterial design, i.e., starting with the modulation of differentiable anisotropic property distribution, which then enables the construction of multiscale geometry, and is scalable to design a variety of functional stochastic metamaterials.

For geometric construction, the modulated spherical harmonic function at each position of the design space serves as the probability density distribution, in order to guide the automatic generation of spinodal structures. Note that spinodal structures are preferable for stochastic configuration with controllable anisotropy to mimic the phase separation process in nature [38]. The main reasons for adopting spinodal structure as building blocks of functional stochastic metamaterials are the balance between geometric disorder and continuity, and its implicit representation that enables parametric control for user interactive design. Unlike the spinodal structure initially proposed by Kumar et al. [31] in which machine learning was adopted to bridge geometric sampling region and target physical property, we directly derived the evolution of spinodal structure via spherical harmonics and theoretically verified its validity.

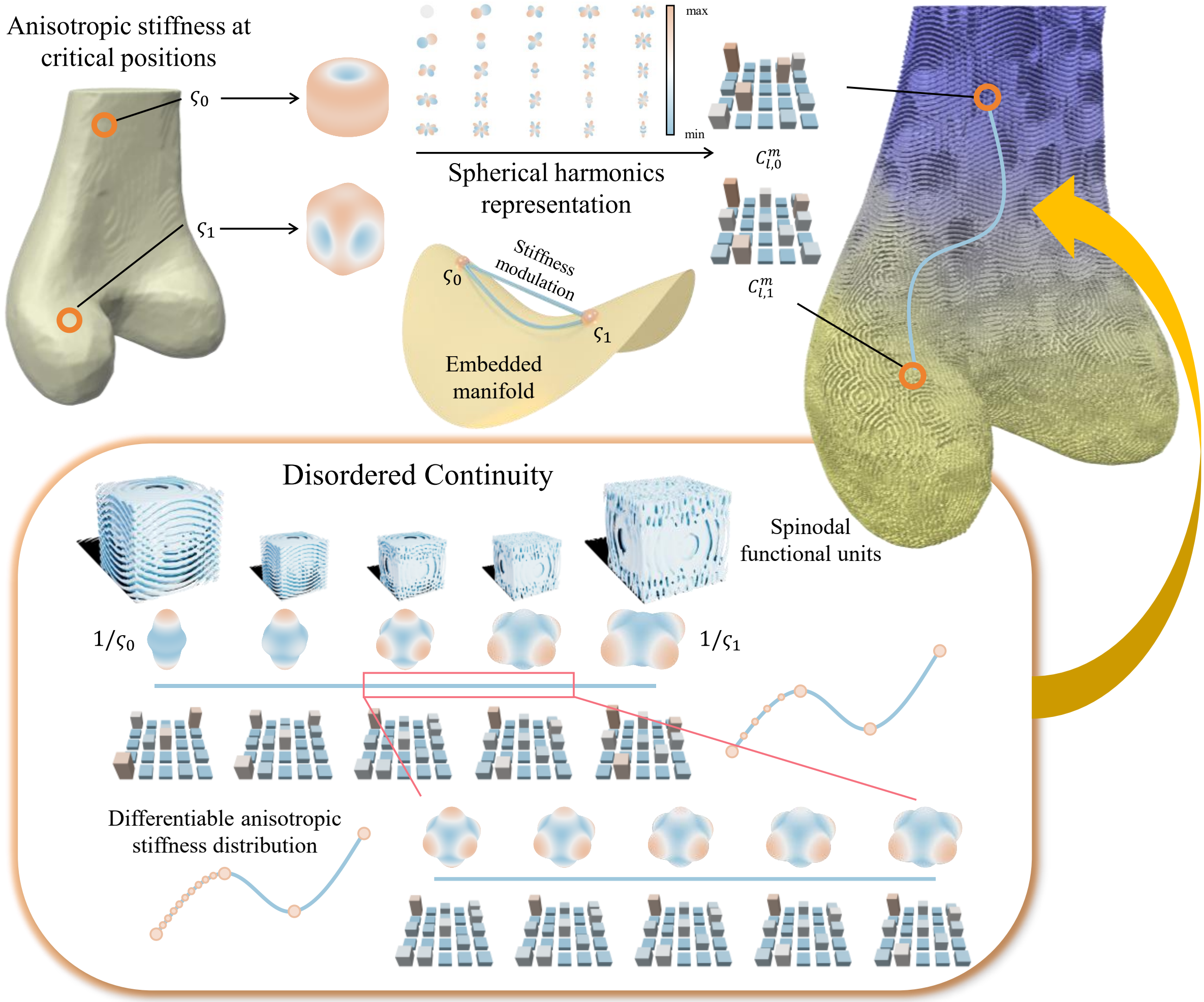

Figure 1. The concept of disordered continuity is enabled by incorporating spherical harmonics to represent and modulate physical property fields (such as stiffness tensor field) for the design of functional stochastic metamaterials. The high dimensional embedding of an original tensor representation of a physical property (e.g., stiffness) cannot be directly interpolated to achieve continuous variation of anisotropy. The spherical harmonics representation of anisotropic property facilitates continuous interpolation of the stiffness tensor (similar to the interpolation of spline curves), which potentiates metamaterial design with arbitrary resolution (termed resolution-independent in this context). Spinodal structure is employed as functional units to recover desired anisotropic stiffness property using controlled sampling scheme for the construction of the entire functional component.

Compared with previously reported studies that mostly strived to develop individual unit cells with specific physical properties, this research developed a global computational scheme to optimize the overall physical property (e.g., stiffness) distribution across the whole structure that facilitates the generation of stochastic metamaterials. Meanwhile, the use of spherical harmonics as intermediate representation established a generalized constitutive relationship between the physical properties and the stochastic geometry. The proposed design method can also be used to construct a variety of functional structures/components that require customized responsive behaviors under external loads, such as the design of artificial femur and meniscus for maintaining sufficient support with reduced stress concentration thanks to enhanced continuity in stiffness. From a much wider view, it may also enable rapid design for applications such as texture encoding, mechanical cloak that requires delicate modulation of stiffness distribution. The evaluation and test of the proposed design method for these applications are discussed in Section 3. The next section will explain the theory of the proposed design method.

## 2. Resolution-Independent Design via Spherical Harmonics

The design process proposed in this paper is illustrated in **Figure 2**, which consists of two stages. Aiming at a specific functional property, the anisotropic physical property, e.g. stiffness tensor at critical positions is characterized and modulated via spherical harmonics in stage 1 (Figure 2(a)). The modulation is independent of the actual geometry that manifests such anisotropic property. By adopting spherical harmonic representation, the stiffness tensor across the entire design space is differentiable and thus can be interpolated to conform to a gridded space with arbitrary resolution. This helps to generate spinodal functional units in stage 2 in accordance with the desired stiffness tensor, so as to form the entire functional components with expected functional property, as shown in Figure 2(b).

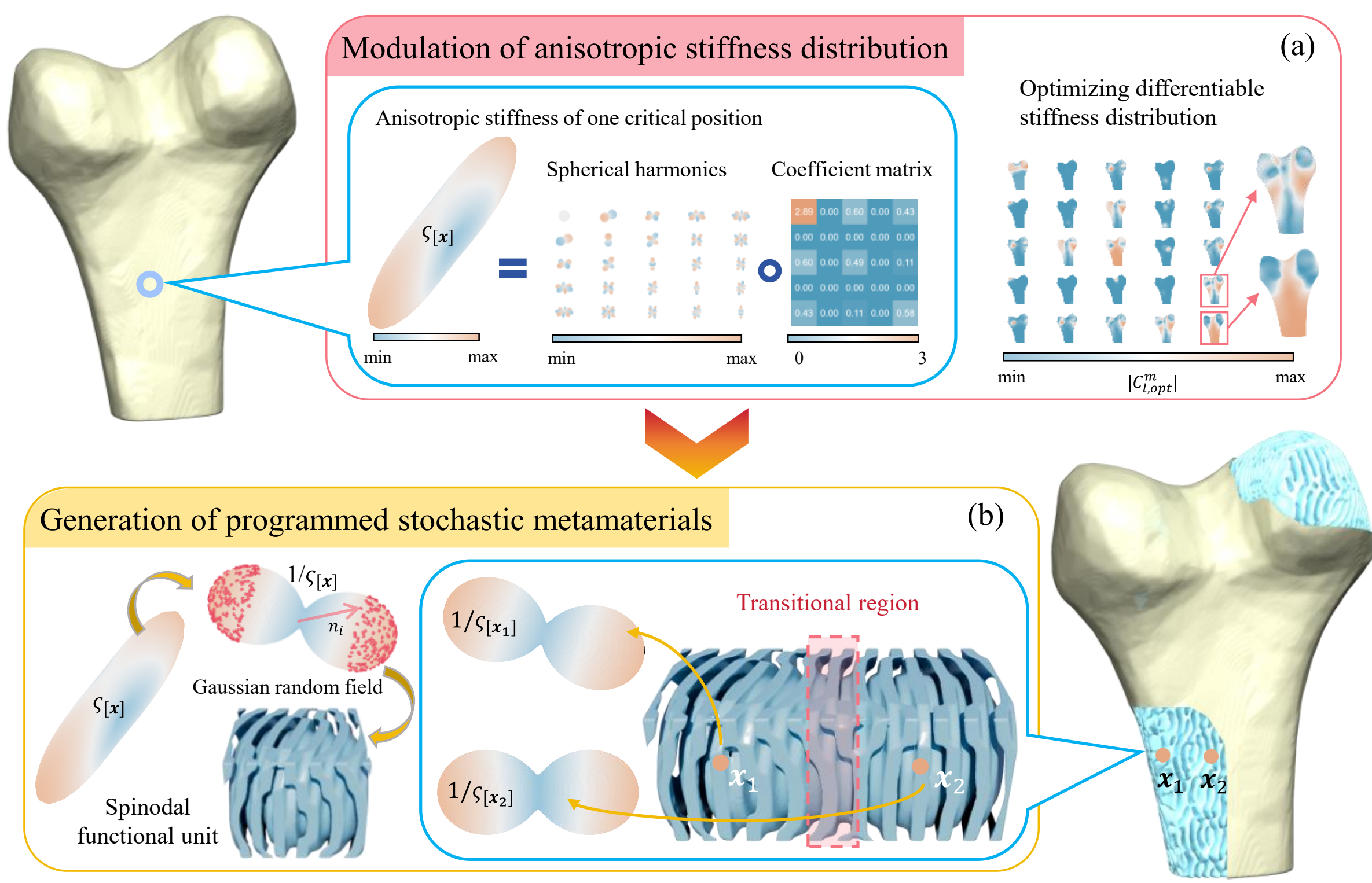

Figure 2. The design process of resolution-independent stochastic metamaterials via spherical harmonics. a) In stage 1, the specified anisotropic stiffness at some critical positions of the design space is converted to spherical function and represented by spherical harmonics (SH), to enable optimization of differential stiffness distribution over the entire design space. b) In stage 2, the optimized spherical function at each position is referred to as probability distribution for generating disordered spinodal functional units, which are used for constructing the stochastic metamaterial for functional components.

### 2.1 Spherical Harmonics-Based Stiffness Representation and Modulation

To achieve the customized mechanical response for a designed component, we set out to modulate the stiffness distribution within the design space, rather than directly generate detailed geometric profile. The physical property stiffness is defined as a measure of a structure's resistance to deformation under external loading. Here, we focus on the mesoscopic perspective of effective stiffness, i.e., the directional effective Young's modulus to delineate the anisotropic stiffness for an infinitesimal element positioned at $\boldsymbol{x}$. This can be represented by a spherical function, which is referred to as stiffness function $\varsigma_{[x]}$ in this paper. Let $l_1$, $l_2$, $l_3$ denote the direction cosines of a unit vector in space with respect to the three principal axes X, Y and Z. The stiffness function to represent the magnitude of the directional Young's modulus can be expressed as:

$$\varsigma_{[\boldsymbol{x}]}(l_1, l_2, l_3) = 1/\sum LS(\boldsymbol{x})L^T \tag{1}$$

where, $L = (l_1^2, l_2^2, l_3^2, l_2 l_3, l_1 l_3, l_1 l_2)$, $S(\boldsymbol{x})$ is the compliance tensor at position $\boldsymbol{x}$.

Therefore, the modulation of the spatial distribution of stiffness function $\varsigma_{[\boldsymbol{x}]}$ is the primary goal to realizing functional design. However, spatial interpolation of position independent function is a crucial but non-trivial operation to realize stiffness modulation, i.e., adjusting local stiffness as well as smoothing the global distribution over the whole structure all require stiffness function $\varsigma_{[\boldsymbol{x}]}$ to be differentiable. Directly interpolating the spherical function at different positions is intractable due to the high dimensionality and its non-Euclidean embedding.

Here we introduce spherical harmonics to characterize stiffness function $\varsigma_{[\boldsymbol{x}]}$ with high fidelity, as shown in Figure 2(a). Spherical harmonics are the extension of Fourier series in spherical domain to encode directional information [39-40], which have been widely applied in computer graphics for real-time rendering. Spherical harmonics are formed by a series of spherical basis functions $Y_l^m \in \mathcal{F}(\widehat{\mathbb{S}}^2, \mathbb{C})$, where $\mathcal{F}(\widehat{\mathbb{S}}^2, \mathbb{C}) := \{f : \widehat{\mathbb{S}}^2 \to \mathbb{C}\}$. Similar to Fourier transform, spherical harmonics $Y_l^m$ can represent arbitrary directional information/function through a set of orthogonal basis functions $f \in \mathcal{F}(\widehat{\mathbb{S}}^2, \mathbb{C})$. This implies that any rational stiffness $\varsigma_{[\boldsymbol{x}]}$ can be expressed as a linear combination of a finite number of spherical harmonics:

$$\varsigma_{[\boldsymbol{x}]}(\theta, \varphi) = \sum_{l=0}^{\infty} \sum_{m=-l}^{l} C_l^m Y_l^m (\theta, \varphi) \tag{2}$$

where, the coefficient $C_l^m$ is obtained through the following integration:

$$C_l^m = \int Y_l^{m*}(\theta, \varphi) \varsigma_{[x]}(\theta, \varphi)\, \mathrm{d}\Omega \tag{3}$$

where, $Y_l^{m*}(\theta, \varphi)$ is the complex conjugate of $Y_l^m(\theta, \varphi)$.

Theoretically, spherical harmonic functions possess orthogonality, normalization, and rotational symmetry, as elaborated in Supplementary Material S1.2. This indicates that the inner product of two different spherical harmonic functions on the sphere is zero (orthogonality), while the inner product of the same functions is 1 (normalization), which enables any spherical function to be uniquely expressed using coefficient matrix of spherical harmonics. Therefore, the stiffness function $\varsigma_{[x]}(\theta, \varphi) = \sum_{l=1}^{n} \sum_{m=-l}^{l} C_l^m Y_l^m(\theta, \varphi)$ at position $\boldsymbol{x}$ can be represented by its associated coefficient matrix $C_l^m$. Therefore, the stiffness function can be continuously modulated by smoothly interpolating coefficient $C_l^m$ of the spherical harmonics during the optimization process. A small change of the coefficient matrix results in smooth transition of spherical function, which guarantees the stiffness function to be differentiable and can be observed in Supplementary Video S1. Since the basis function $Y_l^m(\theta, \varphi)$ can be precomputed in advance, the computational burden for global optimization of stiffness function $\varsigma_{[x]}$ is remarkably reduced.

Through the above steps, we achieved local and global modulation of stiffness distribution for desired functionality. From designer's perspective, one can interactively specify the stiffness at critical positions by adjusting coefficient $C_l^m$ for continuously control of the spherical function $\varsigma_{[x]}$ to realize local modulation. While for global smoothing, continuous stiffness variation of the entire design space can be achieved by minimizing the quadratic energy:

$$\min_{C_l^m} \sum_{(\boldsymbol{x}_i, \boldsymbol{x}_j) \in E} \left\| C_l^m(\boldsymbol{x}_i) - C_l^m(\boldsymbol{x}_j) \right\|^2 + \sum_{\boldsymbol{x}_c} \| C_l^m(\boldsymbol{x}_c) - C_l^m(\boldsymbol{x}_c)' \|^2 \tag{4}$$

Where, the first term represents the smooth change of the coefficients between two adjacent points $\boldsymbol{x_i}$ and $\boldsymbol{x_j}$ of the same edge, and the second term represents the degree of correspondence between the optimized coefficient $C_l^m(\boldsymbol{x_c})$ and the initial coefficient $C_l^m(\boldsymbol{x_c})'$ at critical position $\boldsymbol{x_c}$.

The optimization process is demonstrated in **Figure 3**(a). Taking the femur model as an example, after specifying the required stiffness at critical positions, the global stiffness distribution can be optimized by adjusting the associated coefficient matrix to ensure spatial continuity and programmed anisotropic property. The initial and optimized coefficient distribution is shown in Figure 3(b) and (c) respectively, arranged in the same pattern as the coefficient matrix. It can be observed that the final optimized coefficient is distributed smoothly without sharp variation in key areas with specified stiffness, which indicated the effectiveness of using spherical harmonics for stiffness modulation.

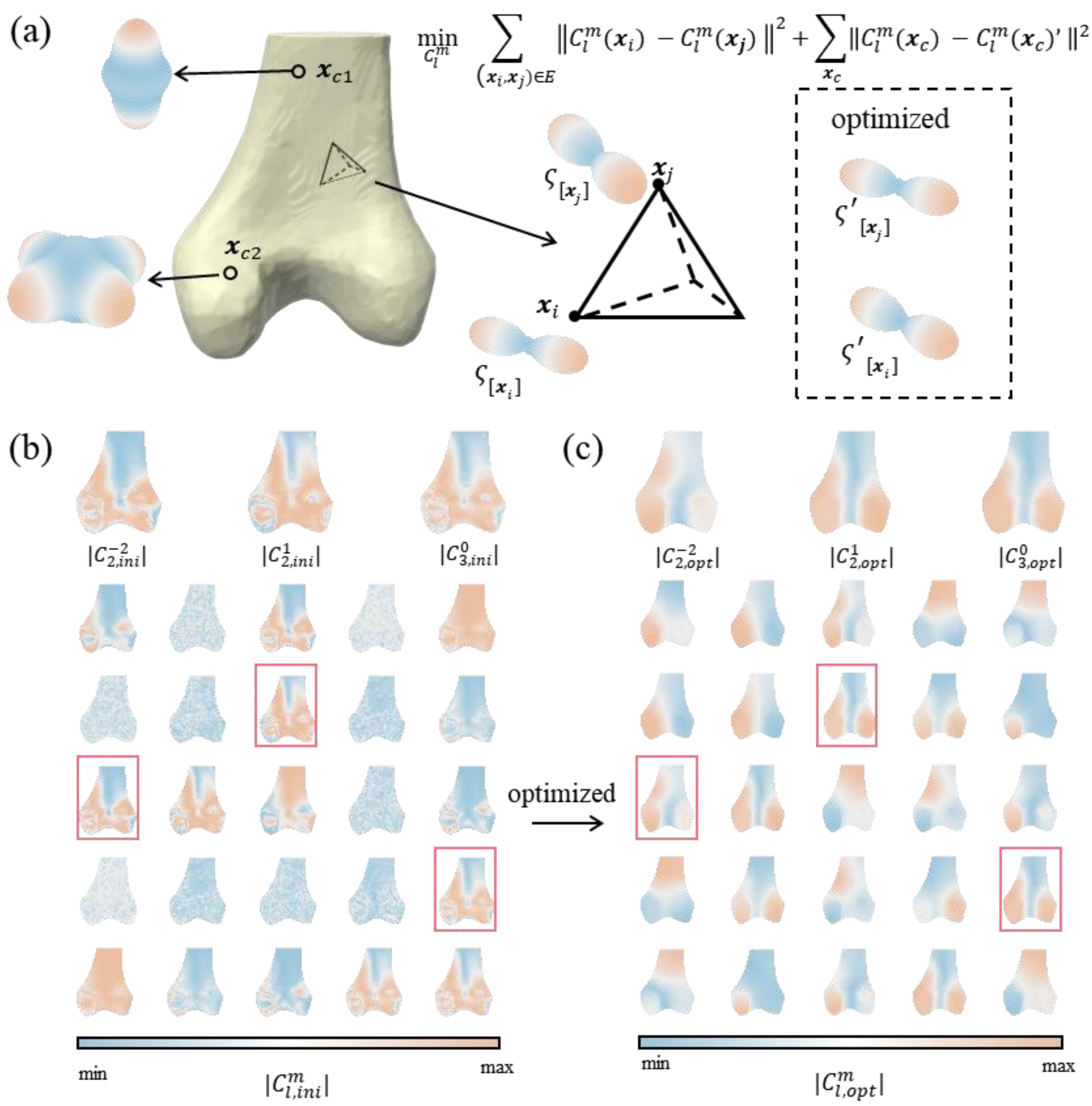

Figure 3. Optimization of anisotropic stiffness distribution via spherical harmonics: a) Objective function is defined for global smoothing of anisotropic stiffness distribution, subject to stiffness defined at critical positions; b, c) Initial and optimized distribution of stiffness property represented by spherical harmonic coefficients in matrix form.

## 2.2 Generating Spinodal Architected Microstructure towards Target Stiffness Distribution

Once the desired stiffness distribution is generated and optimized in the design space, the remaining question is how to design the geometry of the structure that can best achieve the expected property. Inspired by spontaneous mechanism of phase separation in nature, the spinodal decomposition process can be estimated by a sampling process from Gaussian random field (GRF), and directly used to generate stochastic structures [41-42]:

$$\varphi(\boldsymbol{x}) = \sqrt{\frac{2}{N}} \sum_{i=1}^{N} cos(\beta \boldsymbol{n}_i \cdot \boldsymbol{x} + \gamma_i) \quad (5)$$

where $\boldsymbol{n}_i \sim U(S^2)$ denotes a randomly sampled direction from a unit sphere $S^2 = \{\boldsymbol{k} \in \mathbb{R}^3 : \|\boldsymbol{k}\| = \boldsymbol{1}\}$ that determines the direction of the cosine wave function, $\beta$ is a user-defined parameter to control the frequency of wave function, and $\gamma_i \sim \mathcal{U}([0,2\pi))$ represents the phase angle of the wave function.

The spinodal architected microstructure simulated by the transient Cahn-Hilliard equation mentioned above was previously applied to isotropic random structures. Kumar et al. [31] then extended the generation of anisotropic spinodal architected microstructure as functional units through non-uniform probability distribution regulated by conical areas on a unit sphere. By controlling the angle, anisotropic spinodal functional units can be generated and further adjusted:

$$\Lambda(\boldsymbol{x}) = \begin{cases} 1 & \text{if} \quad \varphi(\boldsymbol{x}) \leq \varphi_0 \\ 0 & \text{if} \quad \varphi(\boldsymbol{x}) > \varphi_0 \end{cases} \quad (6)$$

where, threshold $\varphi_0$ was evaluated at the average relative density $\rho = \mathbb{E}[\Lambda]$ of the solid phase.

Our main point of difference from Kumar's approach is the incorporation of stiffness function to directly generate anisotropic spinodal functional units in a generalized manner. The crux here is to establish the relationship between parameters of structural geometry and anisotropic physical property, which was mainly carried out using machine learning in previous approaches. Here we refer to the previously computed stiffness function $\varsigma_{[x]}$ to guide the sampling process:

$$\boldsymbol{n_i} \sim \mathcal{U}(\{\boldsymbol{k} \in f : f > \lambda/\varsigma_{[x]}\}) \tag{7}$$

where $\lambda$ is a user-defined threshold to control the sampling region, initially set to 2/3 by default to avoid overshooting on those non-critical areas. This sampling strategy implies that wave vectors are preferred in the direction of smaller elastic modulus, making the generated spinodal microstructure less rigid accordingly and enabling the acquisition of desired anisotropic stiffness, as illustrated in Figure 2(b) and Supplementary Materials S1.3. The generation process becomes stabilized as the sampling points gradually reach a sufficient amount, as simulated in Supplementary Video S2. The primary reason we adopted spinodal architected microstructure to realize spatially modulated stiffness is for its highly controllable shape and correspondence to the anisotropic physical property. For quantitative analysis, the correlation between the actual stiffness of the generated spinodal unit and the target stiffness function is theoretically derived in Supplementary Materials S1.4. By empirical investigation, we noticed the discrepancy between the actual stiffness function and the target one, which depends on the choice of sampling region $\lambda$ and density threshold $\varphi_0$. In particular, the expansion of the sampling region resulted in a reduction of the stiffness anisotropy of the generated unit, while an increase in the density threshold directly elevated the magnitude of stiffness, as elaborated in Supplementary Materials S1.3. By choosing appropriate values, the generated unit demonstrated sufficient accuracy for recovering stiffness anisotropy to achieve desired global functionality through the assembly of all functional units.

It is noteworthy that the proposed method may be compatible with the incorporation of other forms of stochastic microstructures, leading to more intuitive and flexible design for various functionalities, provided that the correlation between geometry and physical property is clearly established and controllable.

# 3. Results and Discussion

The proposed design method was validated through two phases which demonstrated its potential applications and superiority over traditional methods. Phase one evaluated its ability in designing generalized spinodal microstructures of functional units with high anisotropy of effective stiffness, as a replacement of regular lattice configuration. Such functional units can then be used to construct stochastic metamaterials with uniquely designed responsive behavior, and with continuous variation across the whole component. In phase two, the method was applied to programming stochastically architected metamaterials from scratch with highly modulated stiffness variation. It was proved that such stochastic metamaterials exhibited both softness and hardness portions in controlled anisotropy to enable functional property, which could be potentially utilized in medical implants with accurately assigned property. The durability of the designed component against external loads was also validated through mechanical tests. Meanwhile, the potential application of the proposed method in the design of texture encoded metamaterials was demonstrated. By tailoring different stiffness at different positions, the resulted metamaterials exhibited controlled responses to enable desired function, such as enabling mechanical cloaking as described in Section 3.4.

## 3.1 Quasi-Crystal Spinodal Architected Microstructures

The effectiveness of applying spherical harmonics for stiffness modulation, was first evaluated by testing its scalability in generating spinodal architected microstructures to recover the stiffness of various crystal systems. A crystal system is a point set configuration with unique geometric and physical property, which can be classified by the symmetry of stiffness tensor [43-44]. There are seven crystal systems in total that represent all crystalline materials. For example, the triclinic crystal system has the lowest symmetry with 21 independent variables to determine its stiffness tensor, as shown in **Figure 4**. Here we chose to design the stiffness property of all seven crystal systems according to their different levels of symmetry, including triclinic, monoclinic, orthorhombic, tetragonal, trigonal, hexagonal and cubic systems, which exhibited different profiles of stiffness tensor, as shown in Figure 4. Each crystal system is represented by a typical molecule, such as the triclinic system represented by $CaC_2$. The stiffness tensor of the corresponding molecules had been previously calibrated [45-47]. Accordingly, we generated the equivalent spinodal functional unit to realize the stiffness of each crystal system, which was evaluated by finite element analysis. For simplicity and without loss of generality, the material density of the structure was set to 1 $\mathrm{kg/m^3}$, the Young's modulus to 1 $\mathrm{Pa}$, and the Poisson's ratio to 0.3. The computed functional units as well as their corresponding stiffness values are listed in Figure 4.

It can be observed that the generated functional units with spinodal architected microstructure could realize the anisotropic stiffness of the crystal system of very high accuracy. This not only proved the validity for generating functional structures with target stiffness by periodic arrangement of specific units, but also implied an underlying applicable design principle for quasi-crystal structures combined with non-periodic units of disordered continuity, featuring continuous geometrical and physical transition, to extend beyond natural materials.

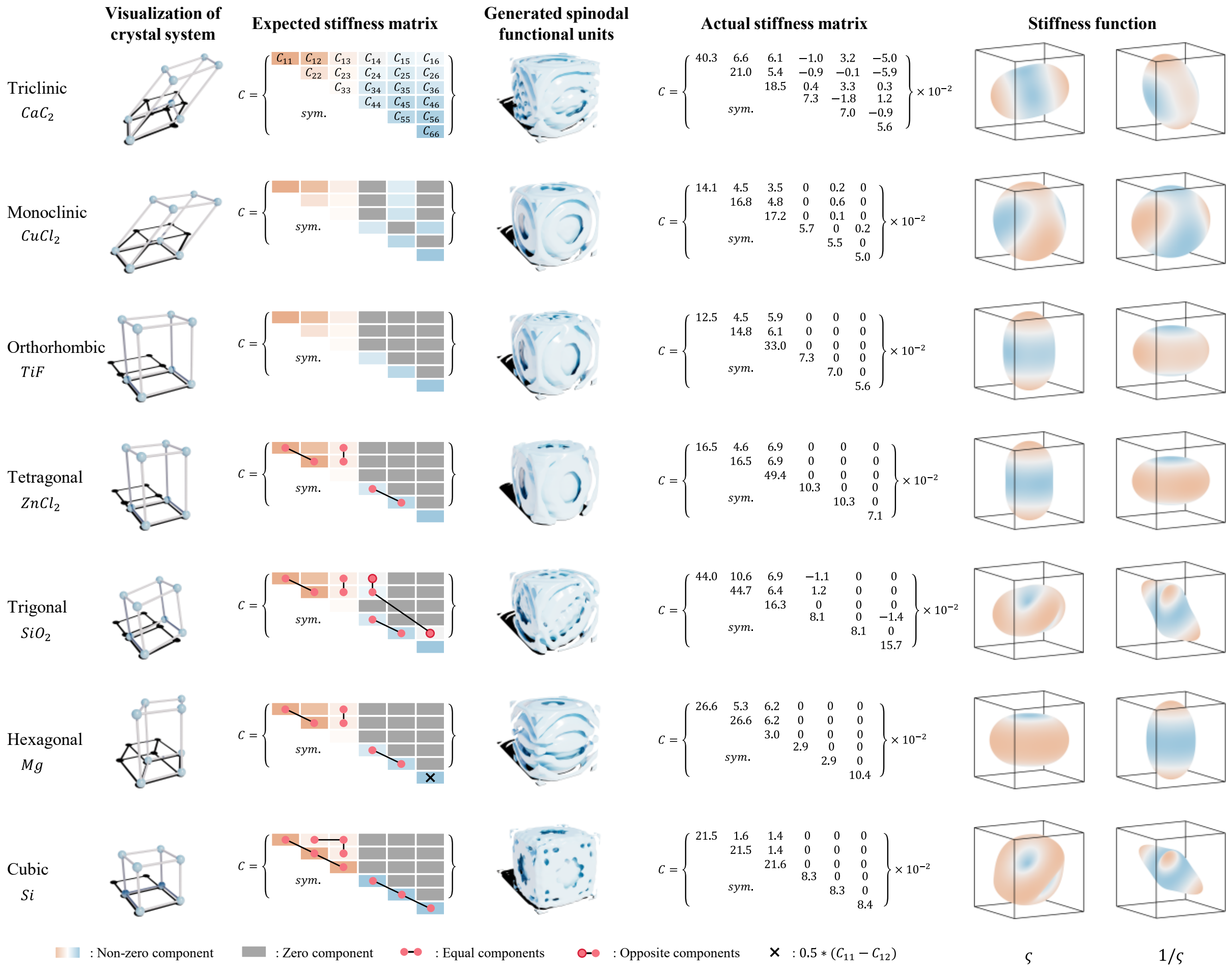

Figure 4. Physical property-driven design for quasi-crystal functional units with highly anisotropic stiffness realization: for all seven crystal systems, corresponding spinodal architected microstructures were generated, whose stiffness was also evaluated with high similarity to that of the desired crystal system.

## 3.2 Design of Biomedical Structures with Tailored Stiffness Distribution

The proposed method was applied to the design of various complex geometries, as shown in Supplementary Material **Figure S1**. The results inaugurated a new physical property-driven design scheme as opposed to previous topology-based methods which required iterative simulation and optimization processes. Our method is a step further to the first principle intuitive design in that, by directly specifying the stiffness at critical control positions (**Figure 5**(b)), the entire structure can be automatically generated with smooth variation in stiffness. This new design principle very much resembles the procedure for designing spline curves and surfaces, but extended for volumetric design, which can be intuitively generated and further refined with a selection of control points.

To demonstrate the potential in biomedical applications, we designed the components of a knee joint, including a femur and a meniscus structure. The loading condition was set up in a simplified way as shown in Figure 5(a). Note that this experiment was carried out specifically for demonstrating the ability of the proposed design method, rather than focusing on the bio-compatibility and functionality validation. In real life, the contacting areas need to be sufficiently soft for shock absorption while the rest of the components still providing adequate rigidity for load bearing purpose.

Previous studies already made remarkable progress in designing spinodal structures by stitching multiple functional units together to program stiffness distribution in fixed resolution, while the continuous transition between adjacent units was mainly achieved through geometric interpolation. In comparison, our method further improved the continuity of stiffness distribution (Figure 5(c)) in transitional interface, rather than merely focusing on geometric continuity, which led to reduced level of stress concentration as compared in Figure 5(e). Precisely, the maximum von Mises stress $\sigma_{Mises}$ as well as the equivalent strain $\varepsilon_{eq}$ were effectively reduced to a great margin when compared with geometrically interpolated spinodal structures, while still preserved designed functionality. This indicated one potential advantage of applying differentiable property distribution in functional structure design, in which the stress concentration can be effectively reduced by modulating smooth transition of physical property over the entire design space.

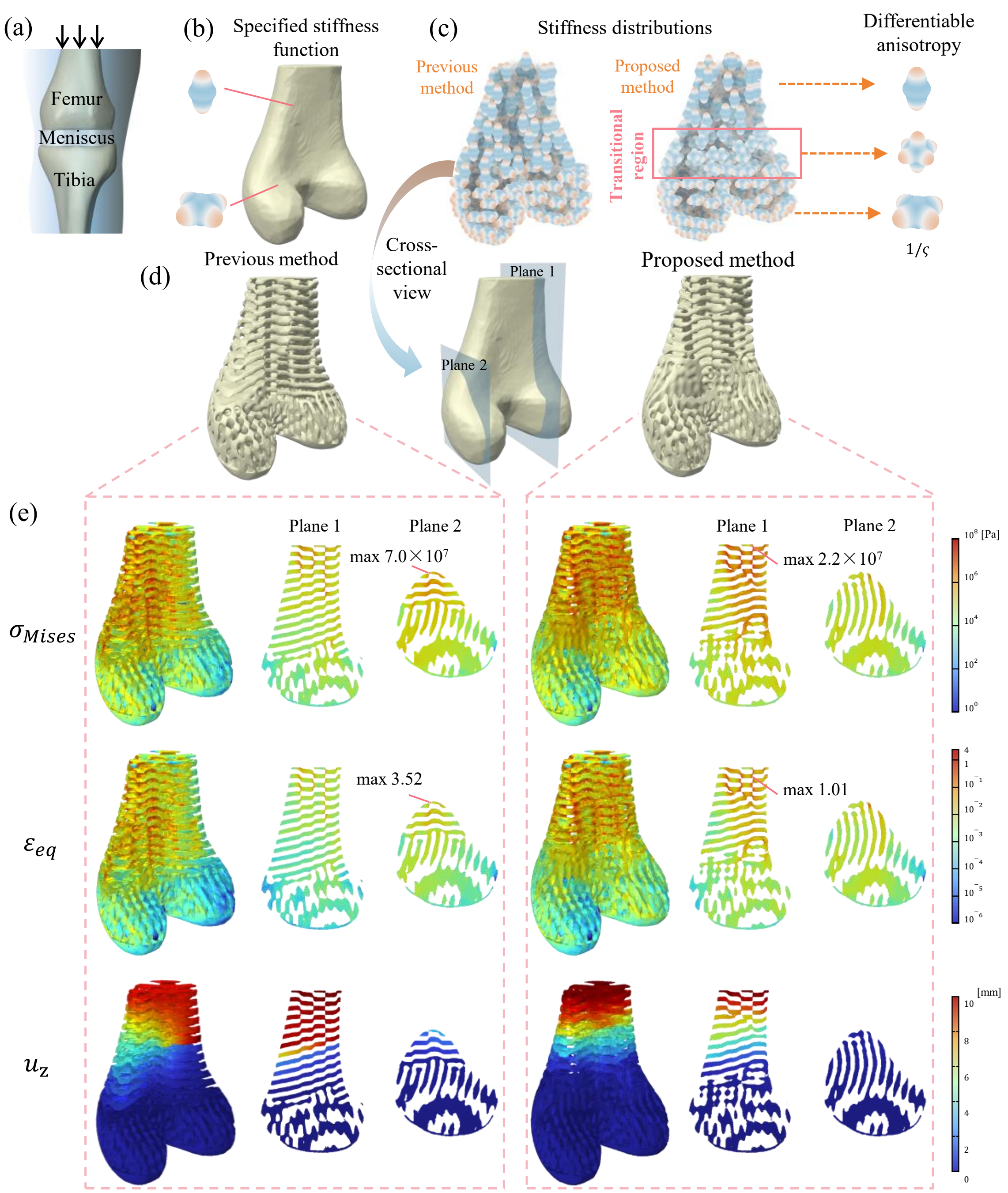

Figure 5. The simulation results of the designed femur structures: a) The simulation setup with boundary condition; b) User-defined stiffness distribution by appointing stiffness function at critical positions; c) Modulated anisotropic stiffness distribution without/with smooth transition using spherical harmonics; d) Generated structures with respect to the stiffness distribution, which are to be sliced for internal check; e) Simulated distribution of von Mises stress, equivalent strain and vertical displacement of the two structures.

In addition, we demonstrated four typical configurations as shown in **Figure 6** for achieving different functionalities. Among them, two types of femur structures were designed with different stiffness distribution along the compression direction. Two meniscus structures were also designed respectively with isotropic and variable stiffness distributions along the compression direction. We used the equivalent strain $\varepsilon_{eq}$ and von-Mises stress $\sigma_{Mises}$ to visualize the responsive behavior from a uniform compression. The final simulation results of the femur and cartilage are shown in Figure 6(c) and (d). By modulating stiffness distribution, the responsive stress, strain and displacement field of the designed structures were altered while maintaining continuity and smoothness of geometry and stiffness.

The designed knee joint structures were fabricated using digital light processing. The resin and printing parameters are elaborated in the Supplementary Material S3. The printed structure shown in Figure 6(e) was tested under the mechanical testing conditions as described in the Supplementary Material S4. From the force-displacement curve in Figure 6(f), it can be observed that the architected structures exhibited different mechanical behaviors as expected. Moreover, both structures shown very strong ductility even with internal cracks occurred during the testing process (indicated by the drop-off spikes of the diagram), which was indeed the effect of continuity in both geometry and stiffness.

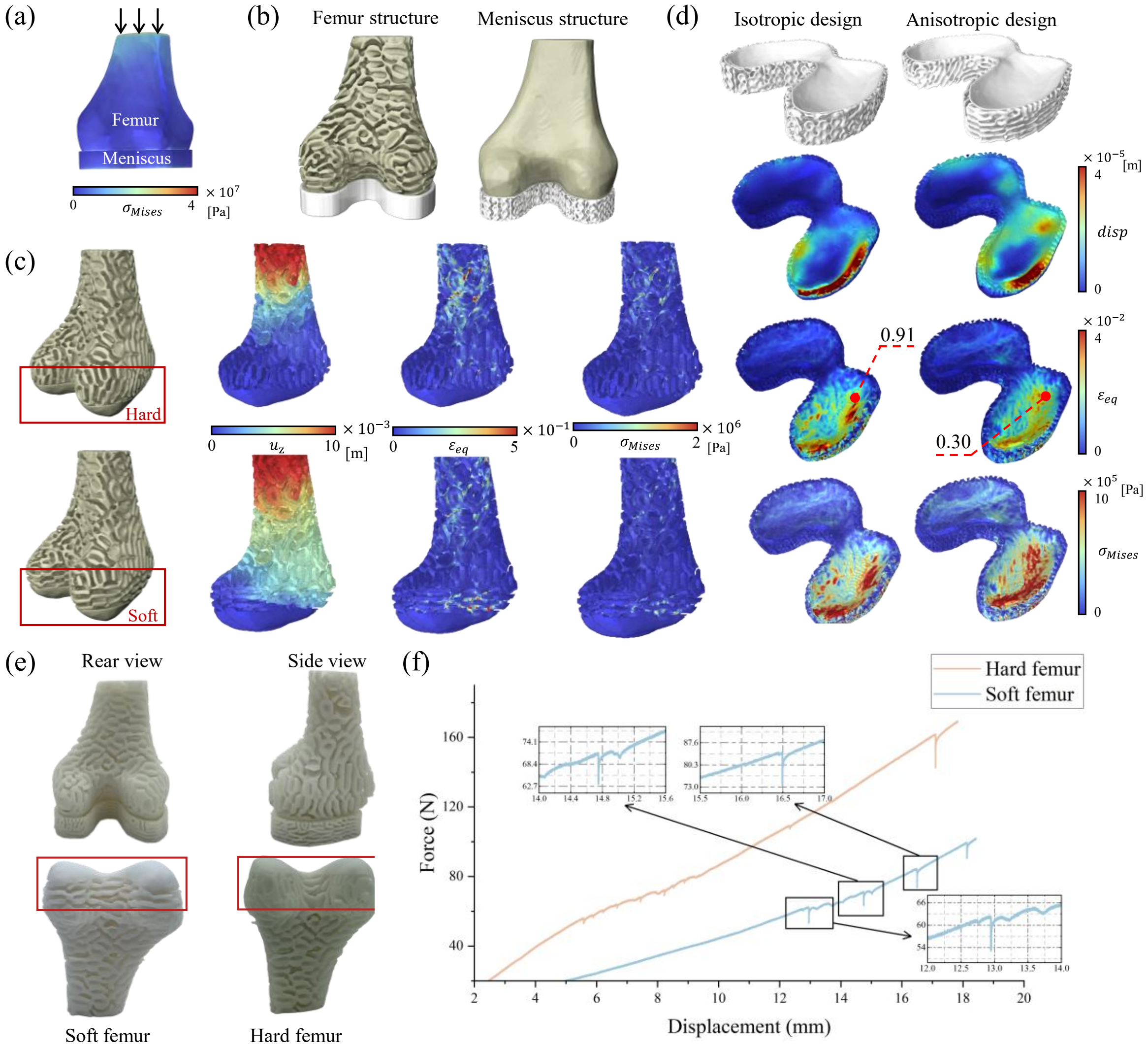

Figure 6. The simulation and physical testing results of a knee joint metamaterial optimized by the proposed method: a) The simulation setup; b) Designed structures of femur and meniscus; c, d) Simulation results of displacement, equivalent strain, and von Mises stress for two different designs of femur structure and meniscus structure; e) Actual printed structures designed with different stiffness distributions; f) The force-displacement diagram of the two femur structures from the compression test.

### 3.3 Design of Textured Metamaterials

The proposed method can also facilitate the design of so-called textured metamaterials by programming polarized deformation under uniaxial load to display informative textures[48]. The texture information can be encoded into the designed structures by precisely modulating the distribution of the anisotropic stiffness. Inspired by previous combinatorial design approach [48], we conducted a preliminary test and observed controlled deformation pattern in a compressed structure formed with different spinodal units, as demonstrated in Supplementary Video S3 and shown in **Figure 7**. In particular, the area with encoded texture '-' was extruded rather than compressed, while the remaining area was compressed with no extrusion due to the modulated stiffness variation. This motivated us to further enhance design flexibility, i.e., high resolution and controllable deformation level of the texture by leveraging differentiable stiffness modulation. To realize this functionality, the encoded area (orange color in Figure 7(a)) should have higher stiffness along compression direction and larger Poisson ratio to achieve prominent extrusion, while the remaining area (blue color) should possess reduced stiffness along compression direction. Specifically, two typical stiffness functions were employed as building blocks to fill into the textured and background space, e.g. the texture of letter 'N' in Figure 7(d) and an emoji texture in Supplementary Material **Figure S9**. Unlike previous combinatorial design approach, our design method potentiates anti-aliasing of the displayed texture by directly smoothing out the stiffness distribution at the texture boundary, while the size of unit cell can be subsequently determined based on the resolution of the texture. This proved the feasibility of the proposed method for textured information encoding and display.

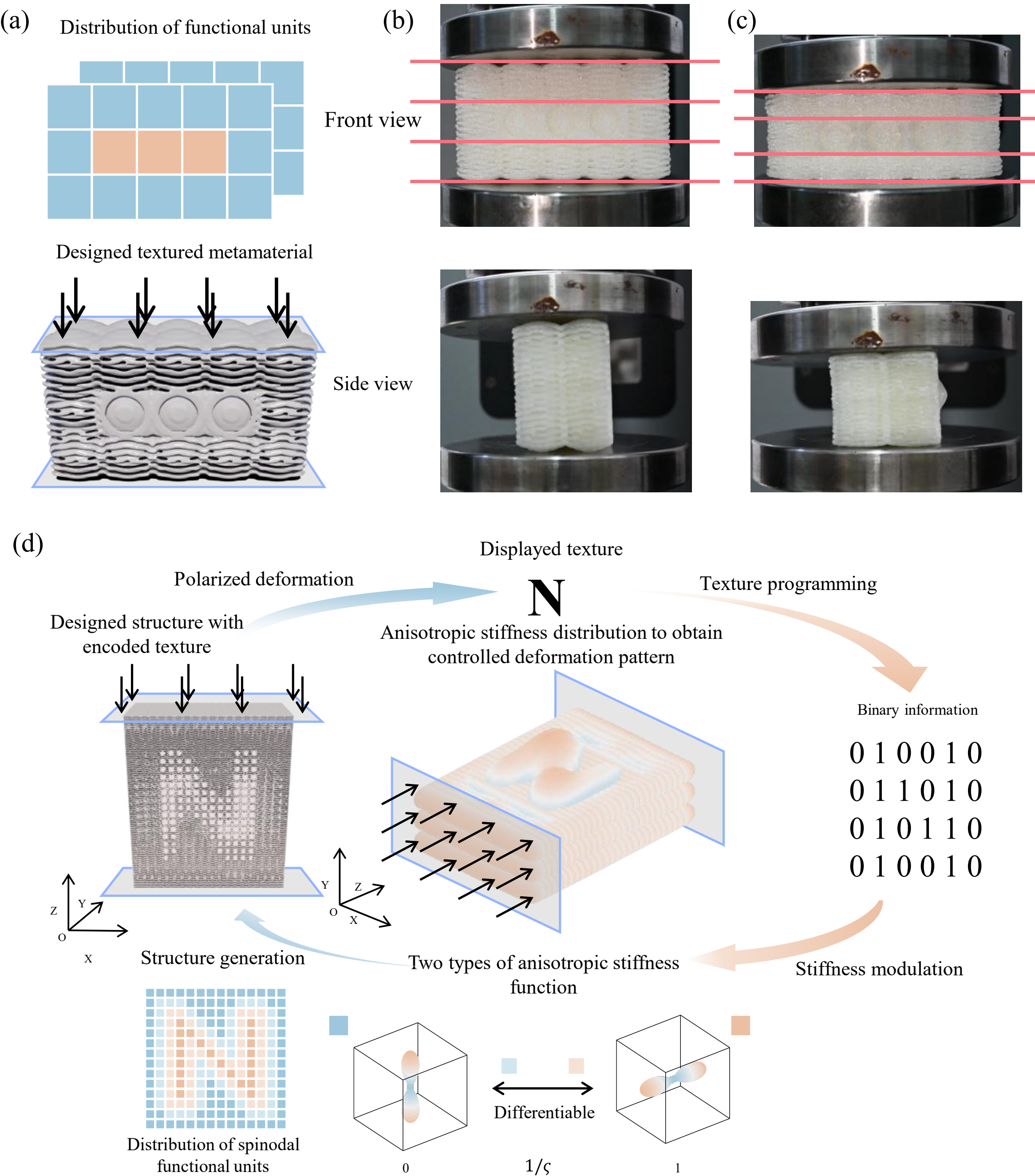

Figure 7. Designing textured metamaterials: a-c) Preliminary testing of a structure composed of two types of spinodal units exhibited controlled deformation to display the encoded texture '-', under uniaxial compression; d) The design process of the texture metamaterial: The texture intended to be displayed under compressive loading is encoded through stiffness modulation to attain polarized deformation. The assignment of a differentiable stiffness distribution, which stems from two extreme types, facilitates high-resolution design and anti-aliasing of the displayed texture, such as letter 'N'.

### 3.4 Design of Mechanical Cloaking Metamaterials

Mechanical cloaking is the concept of manipulating mechanical response such that internal voids/defects are invisible and ineffective from the external responsive behavior under given loading conditions [49]. Similar to optical cloaking, the goal of a mechanical cloak is to control the stress distribution around the internal void to make it less effective to the surrounding region. This can be achieved by iteratively modulating stiffness property distribution at the cloak region such that the external mechanical behavior, e.g. displacement field can fully conceal the internal void.

When designing optical and thermal metamaterials, cloaking effect can be directly achieved by calculating physical property distribution via transformation theory thanks to the form invariance of governing equation. However, this invariance feature is not preserved in elastic mechanics, programming mechanical cloaking effect thus often requires numerical optimization to iteratively refine the anisotropic stiffness distribution based on simulated results.

The overall design process is shown in **Figure 8**(a). Specifically, we took the original homogeneous stiffness distribution $C_{[\boldsymbol{x}]} \equiv C_0$ as initial input, and applied affine transformation including rotation and scaling operation as design variables to the cloak region, such that interior stiffness tensors could be iteratively transformed towards the objective of minimizing the relative displacement error of the surrounding region $O$. The objective function was set as:

$$J = \frac{\int_O \left\| \boldsymbol{u}(\boldsymbol{x}) - \boldsymbol{u}_{ref}(\boldsymbol{x}) \right\|^2 dV}{\int_O \left\| \boldsymbol{u}_{ref}(\boldsymbol{x}) \right\|^2 dV} \tag{8}$$

where, $\boldsymbol{u}(\boldsymbol{x})$ and $\boldsymbol{u}_{ref}(\boldsymbol{x})$ represent the displacements of the optimized and the reference configuration under uniaxial compression.

The optimization process is elaborated in Supplementary Material S4.2. In order to evaluate the cloaking effect, we measured the discrepancy and global error of displacement field based on the following indicator, defined as:

$$\Delta_{\boldsymbol{u}}(\boldsymbol{x}) = \frac{\boldsymbol{u}(\boldsymbol{x}) - \boldsymbol{u}_{ref}(\boldsymbol{x})}{\boldsymbol{u}_{ref}(\boldsymbol{x})},\ \Delta = \sum \Delta_{\boldsymbol{u}}(\boldsymbol{x}) \quad (9)$$

From the simulation results shown in the Figure 8(b), it was found that the maximal displacement discrepancy $\max\left(\Delta_{\boldsymbol{u}}(\boldsymbol{x})\right)$ was theoretically reduced from 16.4% to 6.6%, while the global error $\Delta$ from 171.6% to 19.2%, indicating noticeable cloaking effects as expected.

The main difference of our method from existing mechanical cloak design approach [49] is the capability to smoothly interpolate the optimal stiffness distribution when refined rasterization is provided for microstructure filling, without the need to recompute from scratch. Essentially, assembling functional units of smaller size leads to more refined property distribution and hence closer to the theoretical limit of functionality. By leveraging this capability, the test structures were generated with smaller spinodal units and fabricated by high-resolution digital light processing to preserve the details. Digital image correlation (DIC) was used to evaluate displacement field during the compression test. As depicted in Figure 8(c), the designed cloak structure produced very similar displacement field to the ground truth, while the defect structure exhibited observable difference to be detected. The measured force-displacement curve also indicated similar slope (stiffness along compression direction) between cloak and reference structures, implying strong evidence of cloaking effect. More technical details can be found in Supplementary Materials S3 and S4 and Supplementary Video S4.

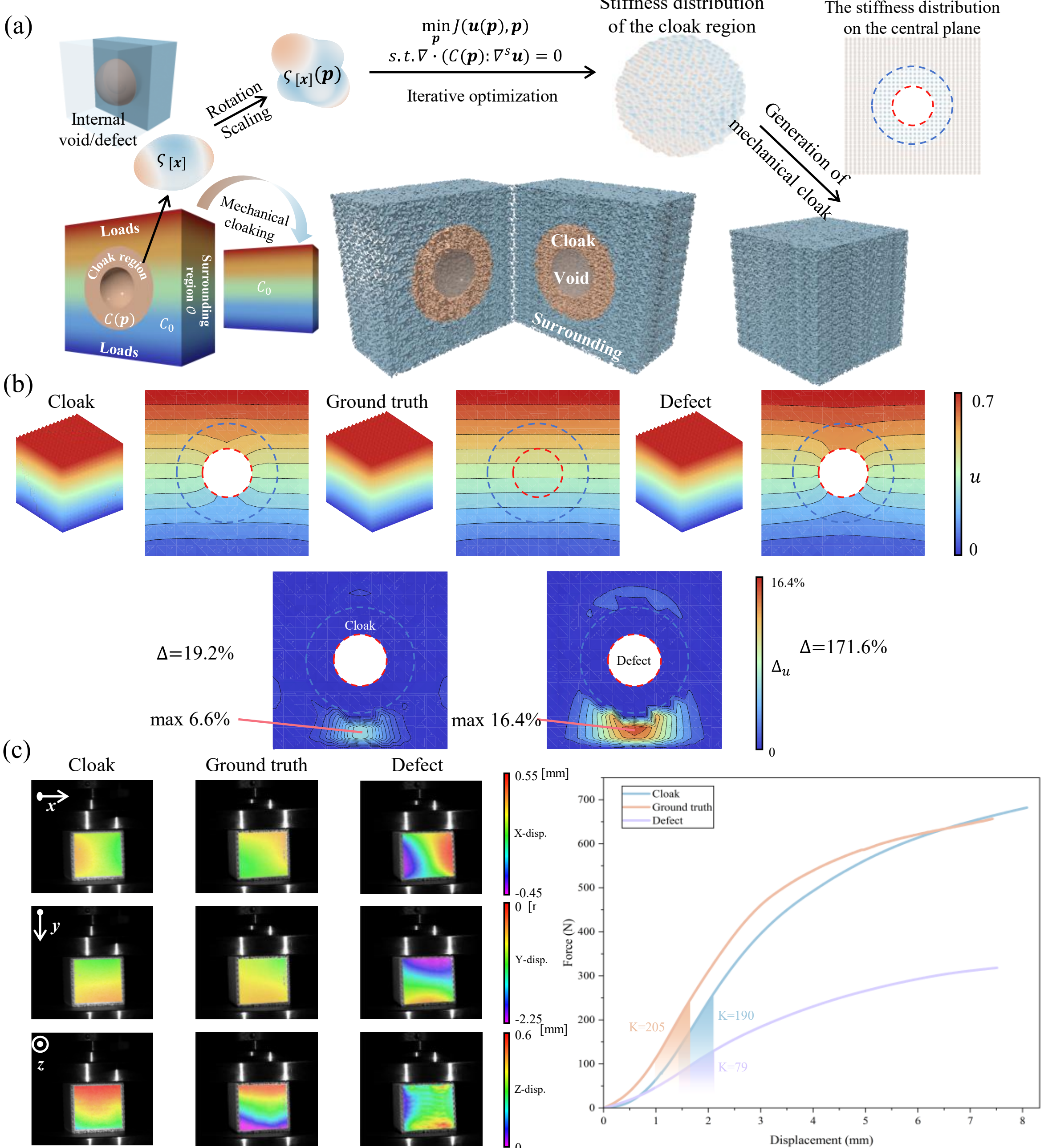

Figure 8. Design of mechanical cloaking metamaterials: a) Mechanical cloaking is programmed by iteratively optimizing the stiffness tensor inside the cloak region, such that the mechanical response of surrounding region was not affected by the internal void. Cloaking metamaterial was generated by assembling spinodal functional units at posteriorly determined resolution, refined to maintain theoretical functionality to a great extent; b) Simulation results of displacement field under uniaxial compression demonstrated a well-established cloaking effect; c) Measured displacement fields and force-displacement diagrams of the actual printed structures, including cloak, ground truth and defect ones, indicating successful cloaking effect.

## 4. Conclusions

This study proposed and proved a physical property driven method for designing multiscale stochastic metamaterials to achieve programmed mechanical responsive behaviors, such as tailored deformation behaviors for texture display and cloaking functionalities. Theoretically different from existing metamaterial design methods of directly assembling functional units under fixed resolutions, the proposed method starts with the modulation of property distribution, e.g., stiffness tensor distribution, over the design space by incorporating spherical harmonics representation to make it spatially differentiable for optimal design, thereby can significantly enhance design flexibility. According to the specified stiffness distribution, spinodal functional units can be directly generated under posteriorly determined resolution to construct the stochastic metamaterial with disordered continuity. The proposed design principle can be directly applied to facilitating crystal-like behaviors and beyond, and can be further extended to designing functional metamaterials with specialized properties, such as artificial knee joint structures with tailored and smoothly transitioned stiffness distribution, textured metamaterials and mechanical cloaking metamaterials with programmed deformation behaviors.

This method has great potential in designing functional stochastic metamaterials. It paved the way to a more intuitive design paradigm for functional metamaterials directly from desired physical property to the realization of final geometry, which is more advantageous over traditional 'Lego-like' design approach. In principle, the method may also be further extended to the design of metamaterials driven by other physical properties (rather than just stiffness), such as thermal and electromagnetic conductivity distribution, targeting at developing a more generalized design paradigm for functional metamaterials.

## Data availability

Data will be made available on request.

## Acknowledgements

This research was funded by the National Natural Science Foundation of China (Ref.: 52175466), and the Outstanding Youth Foundation of Jiangsu Province (Ref: BK20240184).

Supporting Information

**Disordered Continuity: Programming Resolution-Independent Stochastic Metamaterials with Differentiable Anisotropic Property Distribution**

*Canhui Lin* [a,†], *Ke Xu* [a,†], *Chenli Zhou* [a], *Yubin Gao* [a], *Yingguang Li* [a,*]

b. Canhui Lin, Ke Xu, Chenli Zhou, Yubin Gao, Yingguang Li

College of Mechanical and Electrical Engineering, Nanjing University of Aeronautics and Astronautics, Nanjing, China

F-mail: liyingguang@nuaa.edu.cn

† Canhui Lin and Ke Xu contributed equally to this work.

Funding: This research was funded by the National Natural Science Foundation of China (Ref.: 52175466), and the Outstanding Youth Foundation of Jiangsu Province (Ref: BK20240184).

Keywords: stochastic metamaterial, spherical harmonics, stiffness programming, anisotropic property distribution

---

[*] Corresponding author. Email address: liyingguang@nuaa.edu.cn

# S1 Design principles

## S1.1 Programming stiffness distribution for tailored mechanical response

In structural optimization, the distribution of stiffness plays a crucial role in determining the performance of the structure. By programming stiffness distribution, the mechanical response of the structure can be tailored to achieve special functionalities. Towards this target, we developed a numerical optimization method to modulate the anisotropic stiffness of the structure based on the given loading conditions and constraints. The core idea of programming stiffness distribution can be realized in two alternative ways based on different design objectives. One is to specify the stiffness tensor $C_{[x]}$ at critical positions and smoothly interpolate the rest to obtain a differentiable stiffness distribution over the design domain, the other is to directly optimize a global stiffness matrix $K$ with respect to certain target function inside the design domain, such that the entire structure exhibits expected mechanical behavior, e.g. tailored deformation field under external loads. Specifically in this study, the achievement of anisotropic stiffness distribution relies entirely on the geometry of the structure, while the constituent material employed to fabricate such a structure is considered isotropic and homogeneous, and does not play a role in generating the anisotropy.

The global stiffness matrix $K$ of the structure is usually represented as a matrix formed by combining the anisotropic stiffness tensor of each unit element with the design space. We control the response of the entire structure by modifying the stiffness tensor of each element. For achieving a specific responsive deformation field of the structure, the following strain energy is used to form the target function:

$$f(K) = \int_{\Omega} \boldsymbol{u}^T K \boldsymbol{u} \, d\Omega \tag{S10}$$

where, $\boldsymbol{u}$ represents the displacement vector of the structure, $K$ is the global stiffness matrix, and $\Omega$ is the design space. Based on the optimization objective, this quadratic form can be utilized as a key metric for optimizing the global stiffness matrix. The optimized global stiffness matrix is then decomposed into the stiffness tensors of each finite element for subsequent computation.

### S1.2 Characterization and spatial interpolation of anisotropic stiffness distribution

In the main text, we mentioned that any spherical function can be expressed as a linear combination of spherical harmonic basis functions, as shown in the Figure S1. For clarity in expression, we rearrange the layout of the spherical harmonic basis functions to matrix configuration illustrated in Figure S1(b).

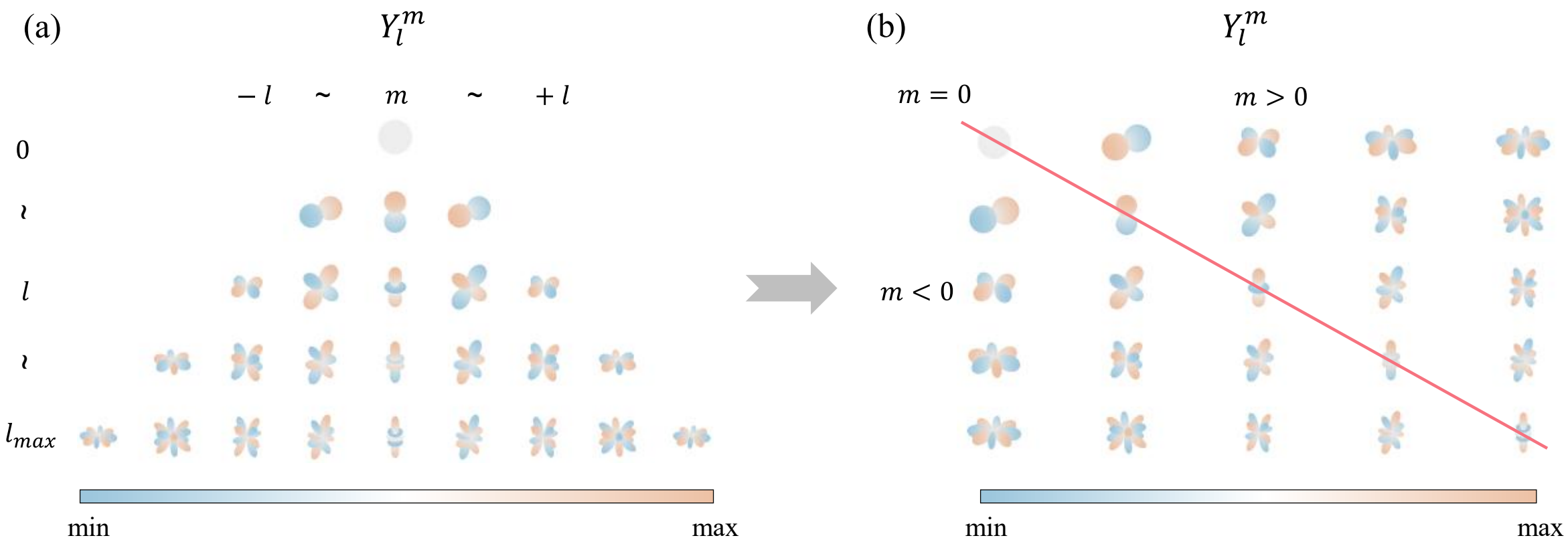

Figure S1. Visualization of spherical harmonics $|Y_l^m(\hat{r})|$ for the first five degrees: a) The first five rows of basis functions correspond to degree $l = 0$ to 4 respectively, and in each row from left to right is its order index $m = -l$ to $l$; b) Rearranged matrix configuration of the spherical harmonics $|Y_l^m(\hat{r})|$.

By applying the orthogonality and normalization properties of associated Legendre polynomials, the following equation of spherical harmonics can be derived:

$$\int Y_l^m(\hat{r}) Y_{l'}^{m'*}(\hat{r}) d\Omega = \delta_{ll'} \delta_{mm'} \tag{S11}$$

where $\delta_{ll'}$ is the Kronecker delta function used to determine whether two quantities are equal. This indicates that the inner product of two different spherical harmonic on the sphere is zero (orthogonality), while the inner product of identical functions is 1 (normalization), which enables any spherical function $f(\theta,\varphi)$ to be uniquely expressed, as shown in Figure S2. This also indicates that any stiffness function $\varsigma_{[x]}(\theta,\varphi)$ can be accurately expressed using the coefficient matrix of spherical harmonics.

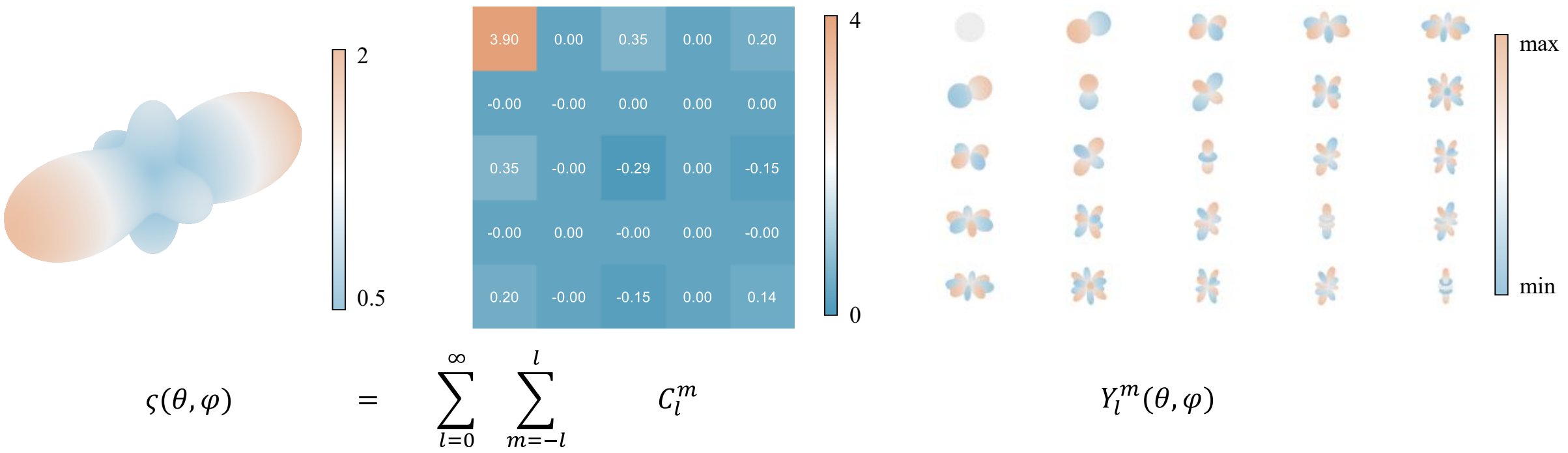

Figure S2. A stiffness function $\varsigma_{[x]}(\theta,\varphi)$ characterized by spherical harmonics $Y_l^m(\theta,\varphi)$, and the corresponding coefficient matrix $C_l^m$.

Given a rotation transformation $R \in SO(3)$, the spherical harmonics can be transformed accordingly:

$$RY_l^m(\theta,\varphi) = \sum_{m'=-l}^{l} D_{m'm}^l(R) \cdot Y_l^{m'}(\theta,\varphi) \tag{S12}$$

where $D_{m'm}^l(R)$ is a complex matrix element, known as the Wigner-D function, which represents the effect of rotation only in the same representation space of degree $l$. This property implies that a rotated spherical function can be expressed by the same degree of spherical harmonics without losing information.

For a specific stiffness functions $\varsigma_{[x]}$ as shown in Figure S3, we found that when using the first five degrees, the representation error by the spherical harmonics is sufficiently small (less than $5 \times 10^{-3}$) from the original function. This is primarily attributed to the physically guaranteed smoothness and symmetry of the stiffness function $\varsigma_{[x]}(\theta, \varphi)$. Here, $l$ in '$\varsigma_{[x]}^{l}(\theta, \varphi)$' represents the highest degree used in the spherical harmonics $Y_l^m(\theta, \varphi)$. It has been empirically proved that when $l = 4$, sufficiently high fidelity is reached for characterizing the stiffness function $\varsigma_{[x]}(\theta, \varphi)$, as expressed below:

$$\varsigma_{[x]}(\theta, \varphi) = \sum_{l=0}^{4} \sum_{m=-l}^{l} C_l^m Y_l^m(\theta, \varphi) \tag{S13}$$

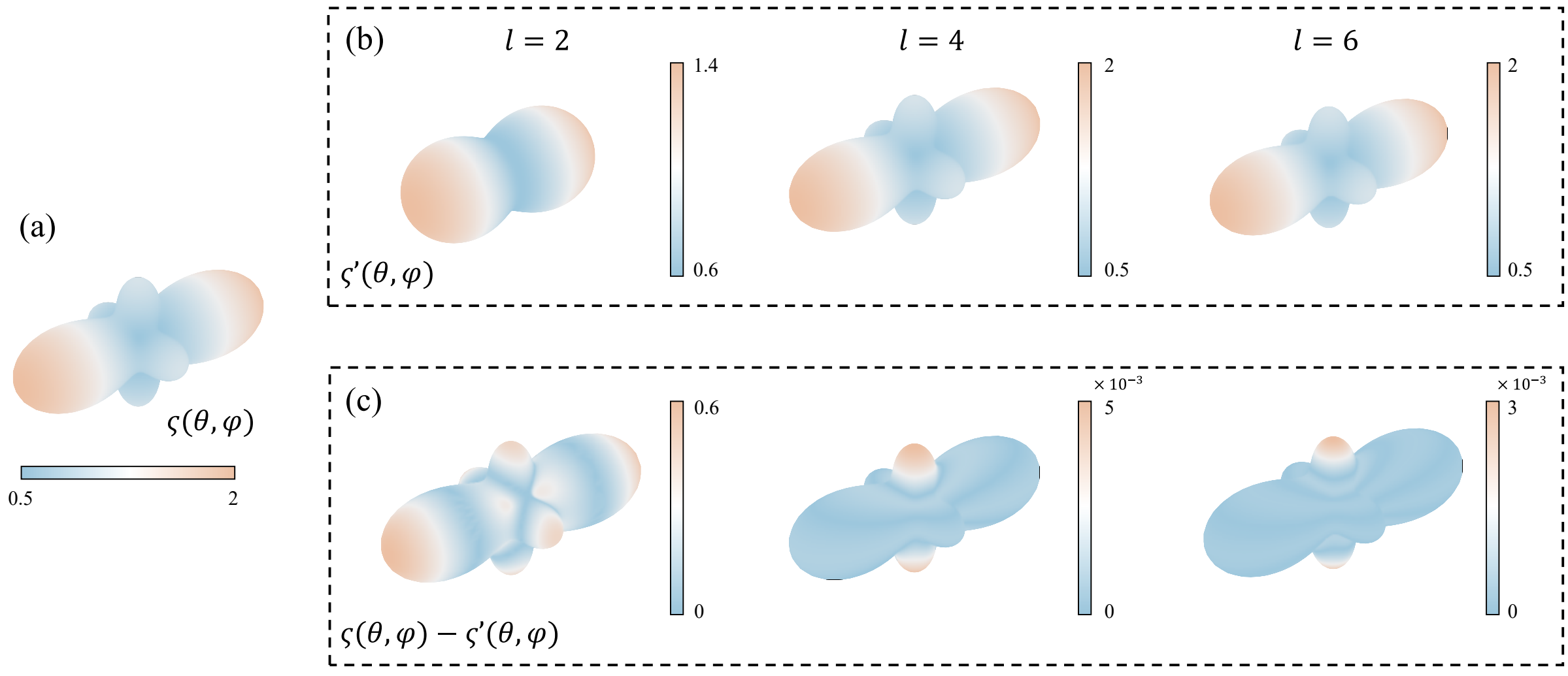

Figure S3. The influence of the degree $l$ on the accuracy of reconstructed spherical function: a) The original spherical function $\varsigma_{[x]}(\theta, \varphi)$; b) Spherical functions $\varsigma_{[x]}^{l}(\theta, \varphi)$ reconstructed with different degree $l$; c) The reconstruction error map with respect to the original spherical function $\varsigma_{[x]}(\theta, \varphi)$.

To demonstrate the correlation between a smoothly changed spherical harmonic coefficient matrix $\mathrm{C}_{\mathrm{l}}^{\mathrm{m}}$ and its represented stiffness function $\varsigma_{[x]}(\theta, \varphi)$, the Supplementary Video S1 is provided for visualization.

### S1.3 Spinodal functional unit generation based on target stiffness function

Despite the resolution-independent characteristic of the proposed method, the generation of the overall functional structure remains dependent on the composition of the spinodal units, whose size can be subsequently specified based on the variation level of the desired stiffness distribution. For each unit, we adopted spinodal microstructures to realize the target anisotropy. As shown in Figure S4(a), the inverse of stiffness function is employed as reference for generation of Gaussian random field (GRF) by sampling random points over the admissible region. During the generation process, a sampling threshold $\lambda$ is used to control the admissible region, in which $N$ points are randomly sampled to generate GRF. A density threshold $\varphi_0$ is introduced to adjust the thickness of the solid phase of spinodal geometry. The influence of these parameters to the generated spinodal structure is demonstrated in Supplementary Video 2. In Figure S4(b), we also demonstrated a set of generated spinodal units based on different values of parameters, and their actual stiffness function as compared to the desired one. Discrepancy between the actual stiffness function and the target one can be observed from the results, although most of them preserve similar anisotropic profiles that are still valid for functional design. A more comprehensive study is worth further investigation regarding how these parameters theoretically influences the final physical property of the spinodal microstructure. Here we rely on empirical choice based on observation, unless otherwise specified, the parameters used in this study remains to be $\lambda = 2/3$ and $\beta = 1.5\pi$.

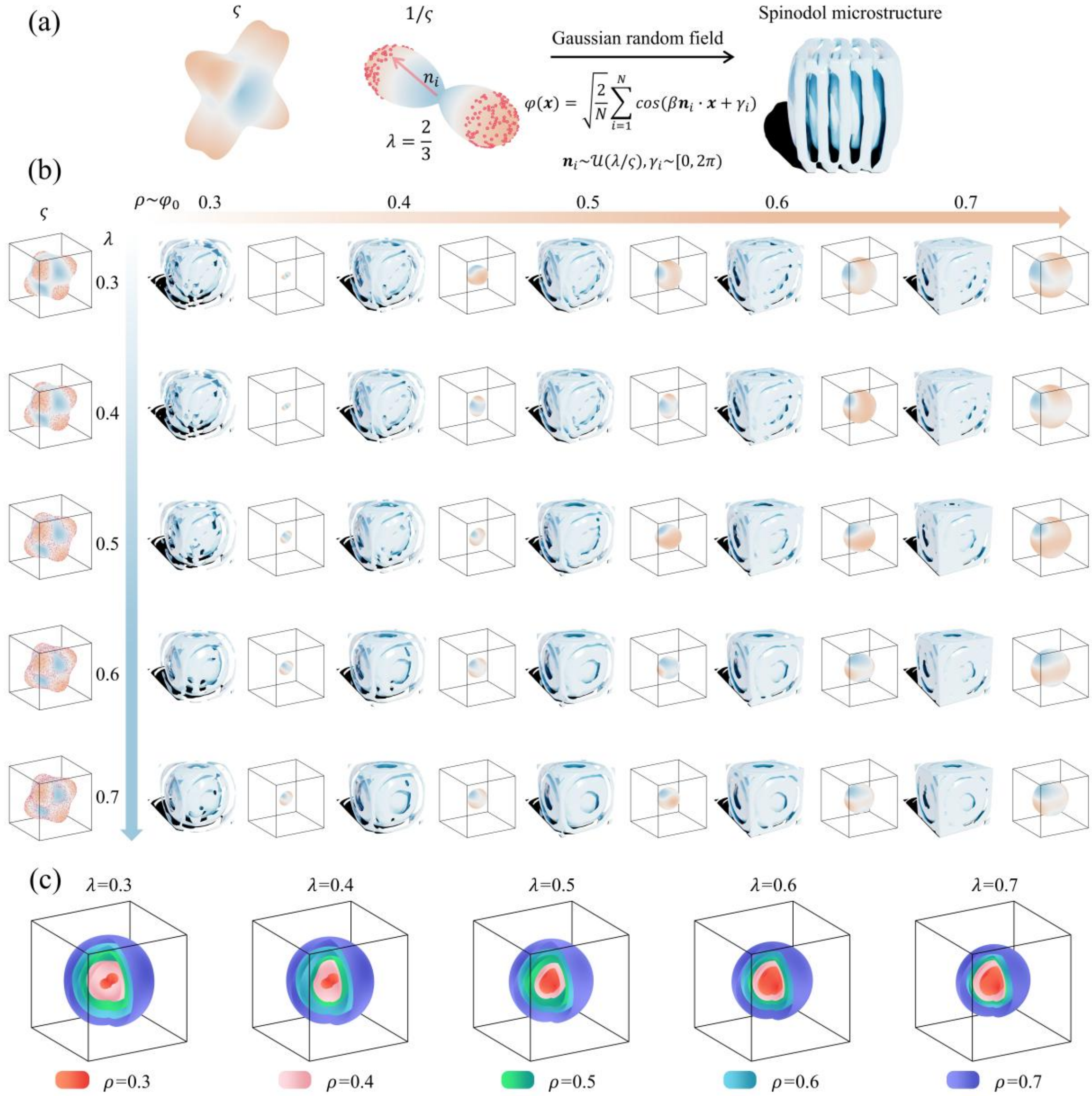

Figure S4. Spinodol functional unit generation process: a) A probability density function is constructed based on stiffness function for random sampling, in order to generate a Gaussian random field and construct spinodol functional units; b and c) The change of the parameters affects the actual stiffness property of the generated spinodal units.

With each GRF of the spinodal microstructures computed in advance, the structure with spatially variable stiffness distribution, as shown in Figure S5, were composed by interpolating GRFs $\varphi_i(\boldsymbol{x}), i = \{1, \cdots, k\}$ in the form of weighted summation for smooth transition:

$$\varphi(\boldsymbol{x}) = \sum_{i=1}^{k} \omega(\boldsymbol{x}, \boldsymbol{x}_i)\varphi_i(\boldsymbol{x}) \tag{S14}$$

where $\omega(\boldsymbol{x},\boldsymbol{x}_i)$ is a radial basis function $\omega(\boldsymbol{x},\boldsymbol{x}_i)$ centered at $\boldsymbol{x}_i$, as shown in Figure S5:

$$\omega(\boldsymbol{x},\boldsymbol{x}_i)=\frac{e^{-\zeta\|x-x_i\|^2}}{\sum_{j=1}^{k}e^{-\zeta\|x-x_j\|^2}} \tag{S15}$$

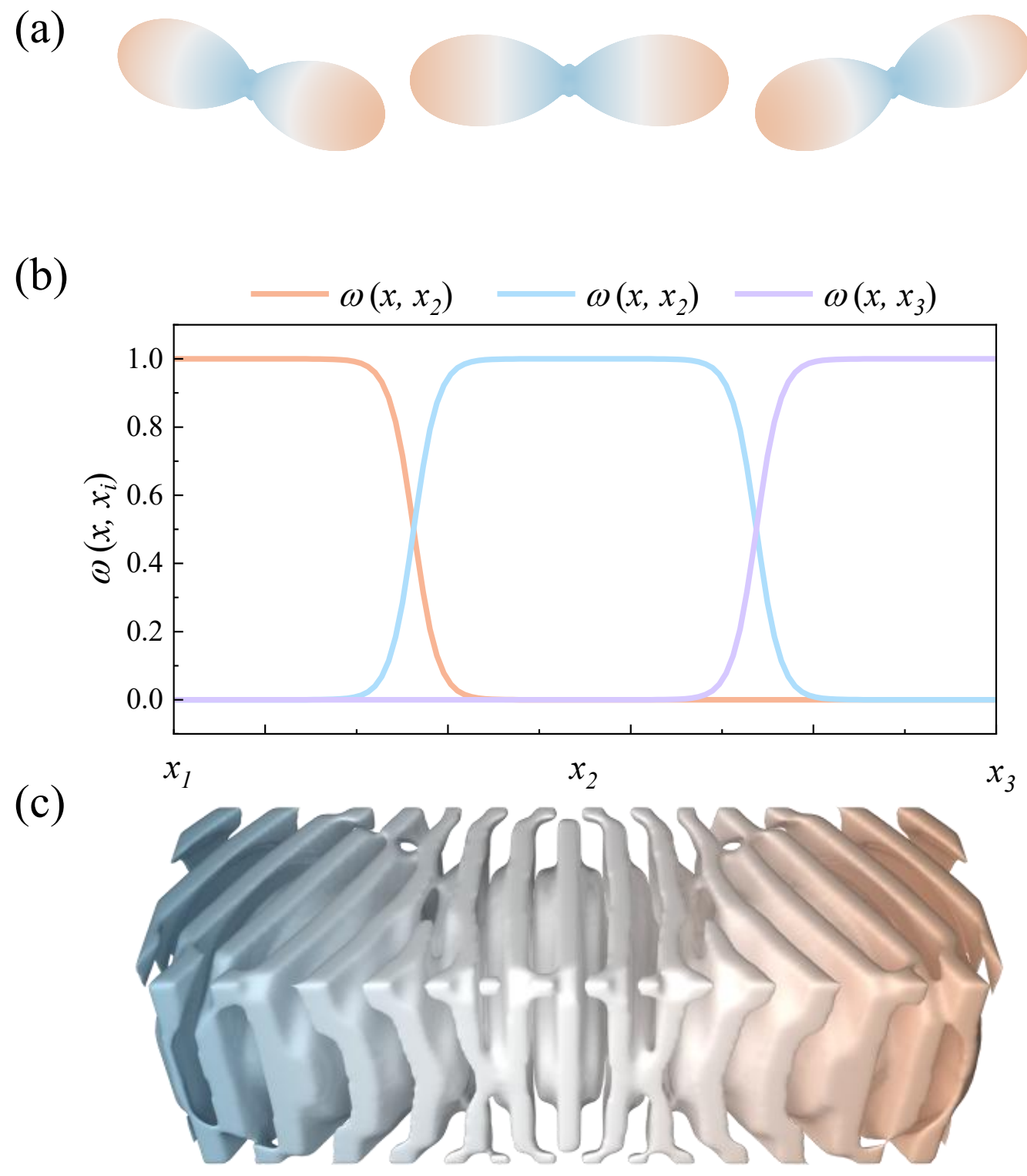

Figure S5. The structure composed of three spinodal functional units with varying stiffness distribution: a) The inverse of three stiffness functions with different orientations ($\pi/6$, $0$, $-\pi$ / 6); b) The radial basis function $\omega(\boldsymbol{x},\boldsymbol{x}_i)$ for geometric transition of spinodal units; c) The final generated structure composed of three spinodal units.

## S1.4 Correlation between target and actual stiffness of the generated functional unit

The spatial arrangement of materials in the spinodal structure is determined by wave vector set $\{\boldsymbol{n}_i\}$ sampled from the spherical function $1/\varsigma_{[x]}$. The statistical distribution of these wave vectors $\boldsymbol{n}_i$ determines the spatial correlation of the structure shape, which in turn affects the equivalent modulus of the overall structure. Equation (7) in main text defines a Gaussian random field whose covariance function is directly controlled by the wave vector set:

$$R(r) = \mathbb{E}[\varphi(x)\varphi(x+r)] = \frac{2}{N}\sum_{i=1}^{N} cos(\boldsymbol{n}_i \cdot r) \tag{S16}$$

Since all wave vectors $\boldsymbol{n}_i$ are unit vectors, the spatial decay of the covariance function $R(\boldsymbol{r})$ reflects the statistical characteristics of the distribution of wave vectors. Here, the volume fraction (percentage of the remaining amount of material) of the selected structure is given by:

$$p = \mathbb{P}[\varphi > \varphi_0] = 1 - \Phi(\varphi_0) \tag{S17}$$

where $\Phi$ is the standard normal distribution function. This generated binary structure exhibits a typical spinodal topology, featuring a disordered yet continuous interface with controllable anisotropy.

Next, we establish the relationship between the structural equivalent elastic modulus $C^{eq}$ and the material distribution through statistical methods. We introduce the two-point correlation function $S_2(\boldsymbol{r})$, which is defined as:

$$S_2(\boldsymbol{r}) = \mathbb{P}[\varphi(\boldsymbol{x}) = 1 \; and \; \varphi(\boldsymbol{x}+\boldsymbol{r}) = 1] \tag{S18}$$

This function characterizes the spatial correlation between material phases, whose form can be indirectly derived from the covariance function of the Gaussian field [50-51]. Since $\varphi(\boldsymbol{x})$ is a Gaussian field, $[\varphi(\boldsymbol{x})\varphi(\boldsymbol{x}+\boldsymbol{r})]$ is a two-dimensional Gaussian vector, and its covariance matrix is:

$$\Sigma = \begin{bmatrix} 1 & R(\boldsymbol{r}) \\ R(\boldsymbol{r}) & 1 \end{bmatrix} \tag{S19}$$

Therefore, $S_2(\boldsymbol{r})$ can be expressed as the two-dimensional normal cumulative distribution function:

$$S_2(\boldsymbol{r}) = \iint_{\varphi_0}^{\infty} \frac{1}{2\pi\sqrt{1-R(\boldsymbol{r})^2}} exp(-\frac{u^2-2R(\boldsymbol{r})uv+v^2}{2(1-R(\boldsymbol{r})^2)})\, dudv \tag{S20}$$

which can be simplified using symbols:

$$S_2(\boldsymbol{r}) = \Phi_2(\varphi_0, \varphi_0;\ R(\boldsymbol{r})) \tag{S21}$$

Here, $\Phi_2$ represents the integral value of the two-dimensional normal distribution in the first quadrant, and the correlation coefficient is $R(\boldsymbol{r})$.

To describe the equivalent elastic modulus of the spinodal unit generated by GRF, we refer to the theoretical framework of statistical mechanics proposed by Torquato et al [50]. This theory proves that for a random medium composed of two-phase materials, its overall mechanical response not only depends on the elastic moduli and volume fractions of each phase, but more importantly, relies on the spatial distribution statistics of each phase, especially the two-point correlation function $S_2(\boldsymbol{r})$, representing the joint probability of "material relative to material phase" in the structure, reflecting the spatial correlation of the structure. Torquato derived the approximate expression of the equivalent stiffness tensor in the general form:

$$C_{ijkl}^{eq} = C_{ijkl}^{(holes)} + (C_{ijmn}^{(solids)} - C_{ijmn}^{(holes)}) \int_{\Omega} \Gamma_{mnkl}\,(\boldsymbol{r})[S_2(\boldsymbol{r}) - p^2]dr \tag{S22}$$

where $C_{ijkl}^{(holes)}$ and $C_{ijmn}^{(solids)}$ are the constitutive stiffness tensors of the two-phase material. $p$ is the volume fraction of solid phase, $\Gamma_{mnkl}(\boldsymbol{r})$ is the response kernel function, which depends on the reference material and loading conditions. $S_2(\boldsymbol{r}) - p^2$ represents the deviation of the two-point correlation function from the expected value under the assumption of spatial independence, quantifying the degree of spatial correlation or dependence in the material at a given distance $\boldsymbol{r}$.

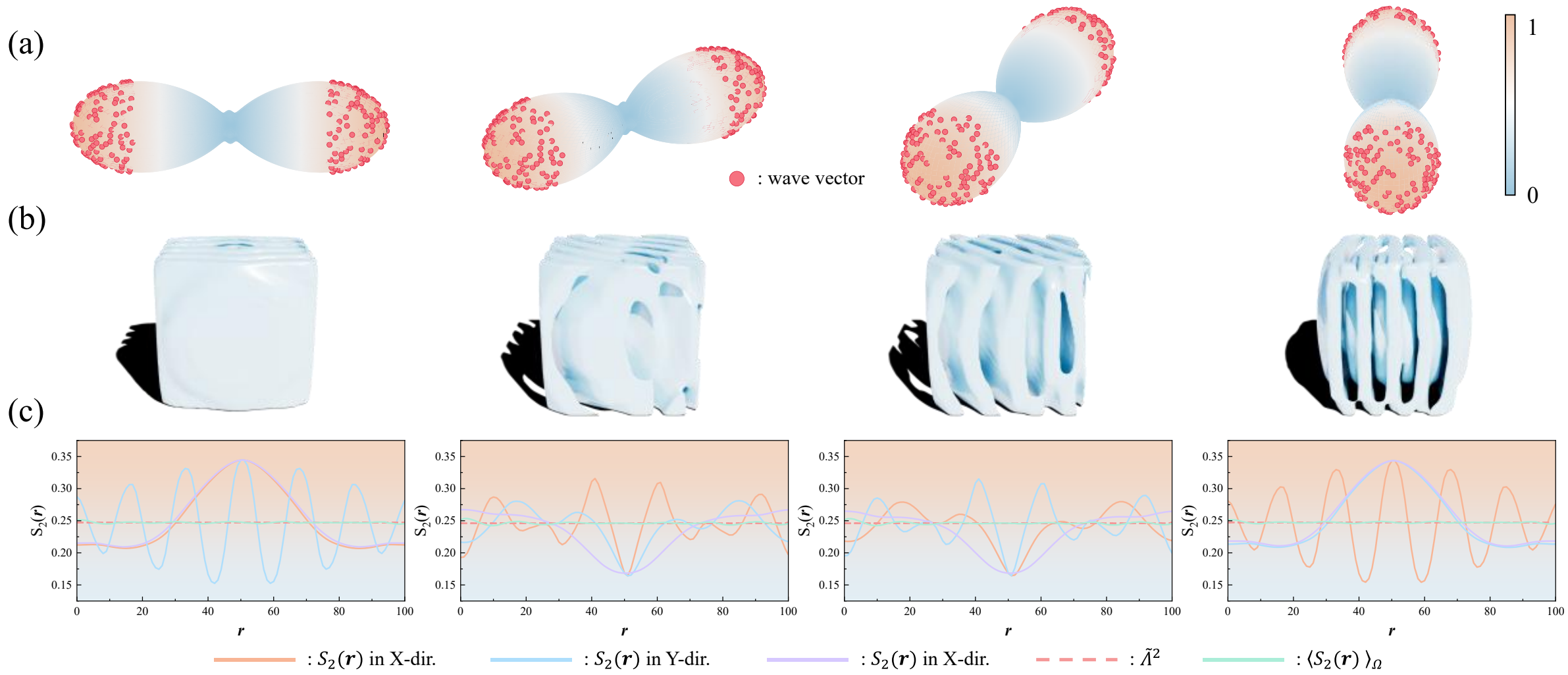

Figure S6. The stiffness property of spinodal unit can be analyzed using the two-point autocorrelation function: a) The inverse of stiffness function $\varsigma_{[\boldsymbol{x}]}$ $(\theta = 0;\ \theta = \pi/6;\ \theta = \pi/3;\ \theta = \pi/2)$ for generating b) the corresponding spinodal units; c) The diagrams of two-point correlation functions $S_2(\boldsymbol{r})$ in the X, Y, Z directions, the relative volume fractions $\tilde{\Lambda}^2$ and the average two-point correlation functions $\langle S_2(\boldsymbol{r}) \rangle_\Omega$.

The equivalent modulus of the structure is influenced by several factors, including the relative material distribution, the degree of anisotropy, connectivity, and long-range correlations within the structure.

- If $S_2(\boldsymbol{r})$ decays slowly in a certain direction, it implies that the material in this direction exhibits strong continuity, often resulting in a higher modulus in that direction.
- If $S_2(\boldsymbol{r})$ varies significantly in multiple directions, the structure is considered anisotropic, as the material properties differ across different directions.
- If $S_2(\boldsymbol{r})$ is disordered, decaying rapidly to the average volume fractions $\tilde{\Lambda}^2$, it indicates reduced modulus along that direction, reflecting weaker material properties.

The generated spinodal units shown in Figure S6(b) can be analyzed as an example. The corresponding spherical function $\varsigma_{[\boldsymbol{x}]}$ of each unit is shown in Figure S6(a). The size of each cubic cell is 100*100*100. The control point of the stiffness function is the cubic centroid. It can be observed that the direction of the spherical function has a significant influence on the actual stiffness of the generated structure. Taking the first case as an example, the spherical function is distributed along the X direction, and the computed two-point correlation function indicates that the $S_2(\boldsymbol{r})$ decays much slower in the Z and Y directions, implying higher modulus in Z and Y directions. In contrast, the YZ direction has a strong oscillation that leads to lower modulus in the YZ direction. Meanwhile, the green lines in the four cases represent the overall radial average $\langle S_2(\boldsymbol{r})\rangle_{\Omega}$, which is slowly decaying to the value of the square of volume fraction, indicating disorder of the structure.

## S2 Miscellaneous design process for each case study

For a particular functional component made of stochastic metamaterials, the anisotropic stiffness property at critical positions is first specified/optimized towards certain mechanical behaviors. Then, for arbitrary positions inside the design space, its associated stiffness property can be realized by spinodal functional unit at specified resolution, as shown in Figure S7. Each spinodal functional unit is initially defined by GRF $\varphi_i(\boldsymbol{x})$, which can be stitched together with radial bases function applied to the interface. The final stochastic metamaterial is extracted in voxel representation and converted to tetrahedron mesh using Marching Cubes algorithm[52-53]. In this way the actual effective stiffness in the scale of functional unit could be evaluated by applying six loading conditions (three axial and three shear loads in orthogonal directions) in finite element analysis.

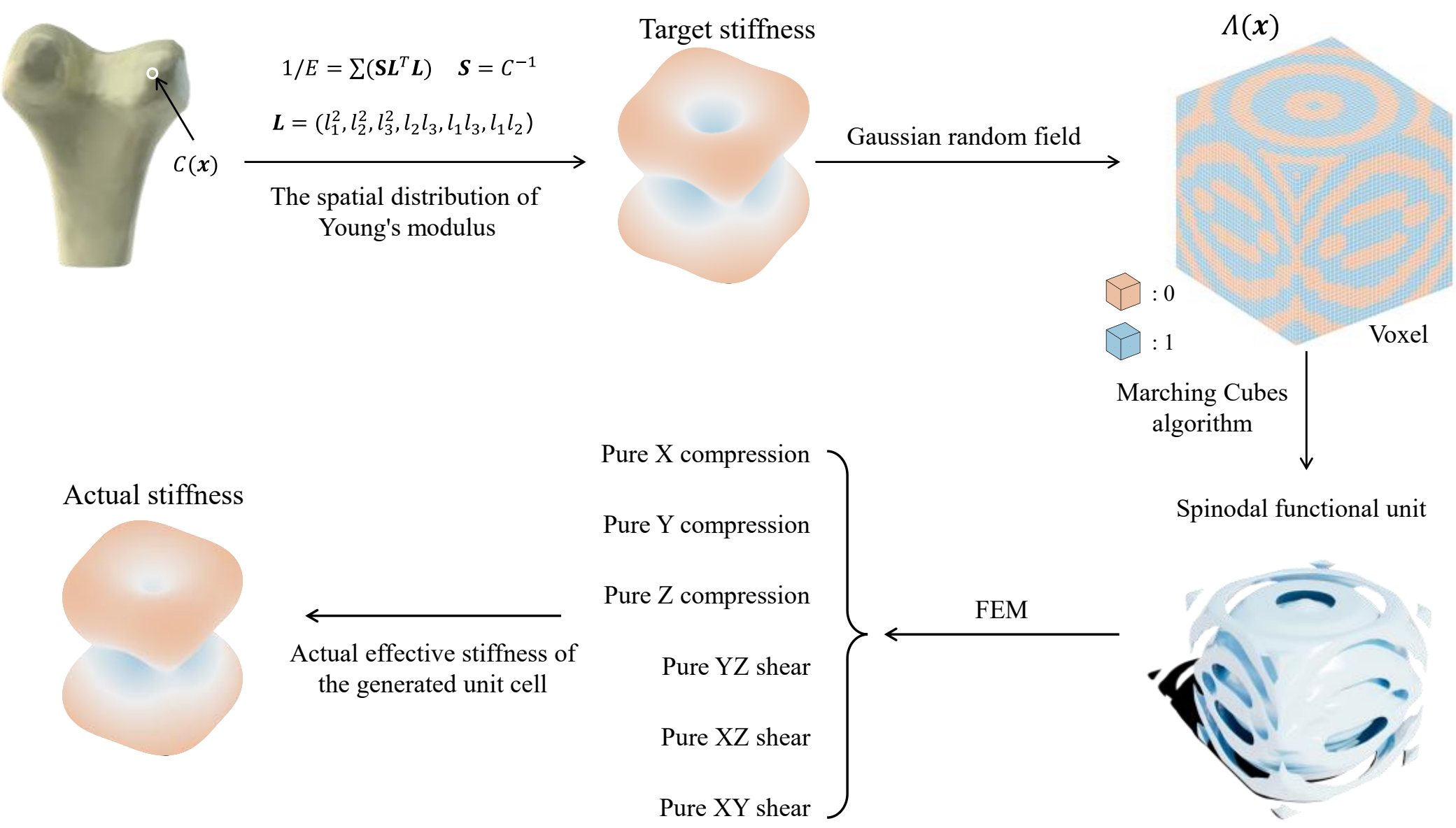

Figure S7. Generation and evaluation of spinodal functional units.

Our method is capable of designing various functional components with complex geometry. These designed structures were composed of continuous varying spinodal units with unique anisotropic stiffness property, as shown in Figure S8.

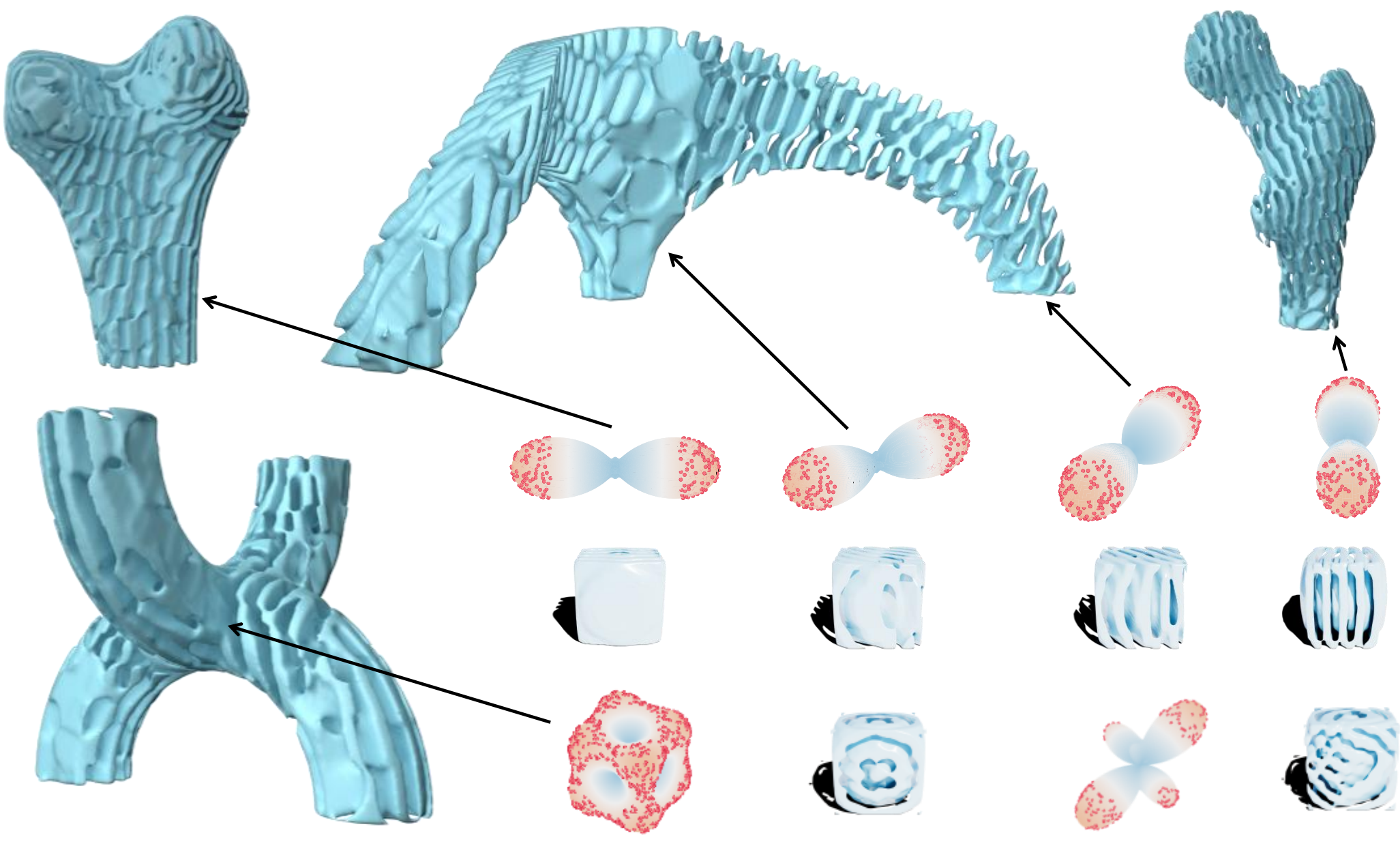

Figure S8. The physical property driven design method used to generate example functional components with differentiable anisotropic stiffness distribution.

### S2.1 Generation of quasi-crystal spinodal functional units

Targeting at the stiffness tensor of typical crystalline lattice, the proposed method can be used to generate quasi-crystal spinodal functional units covering the entire crystal system, providing an alternative way to achieve lattice-like property with disordered geometry. Each functional unit is designed in size of $100*100*100$ voxels. A unit displacement in each direction is applied to evaluate the elastic modulus. The stress and strain under six different conditions are calculated to obtain the stiffness tensor corresponding to each unit. The design parameters are selected as $\lambda = 2/3$, $\beta = 0.7\pi$.

### S2.2 Design of biomedical structures with tailored stiffness distribution

This case study demonstrates the capability of designing structures with tailored stiffness distribution by directly specifying desired value at critical positions. The size of the femoral model is 230*190*290 mm. The simulation conditions is described in the main text. The simulation is conducted by applying a uniform displacement along the z-axis, and obtain the corresponding stress and strain distribution. We found that when designed with smoothly interpolated stiffness distribution, the level of stress and strain concentration at structural interface can be largely alleviated, by observing two cross-sectional planes where the maximum stress/strain occurs. The material was set with density to 1200 $\mathrm{kg/m^3}$, Young's modulus to 25 MPa, and Poisson's ratio to 0.37.

### S2.3 Design of textured metamaterials

Textured metamaterials can be designed by assembling typical functional units with different compressive responses. The corresponding spinodal structure parameters are generated with parameters $\lambda = 2/3$, $\beta = 3\pi$ to achieve high distinction of anisotropic property. Originally, the resolution of displayed texture is in accordance with the structural resolution. The differentiable property distribution achieved via our method enables higher design resolution and potentiates anti-aliasing effects to the texture, realized by encoding the textured information in variable scales with smoother property transition on the boundary. As shown in Figure S9, the Emoji texture is converted into a binary matrix represented by 0 and 1. Unlike conventional design methods that directly assign proper units to the binary matrix, this study achieves anti-aliasing for the metamaterial by further interpolating the stiffness property at the texture boundaries.

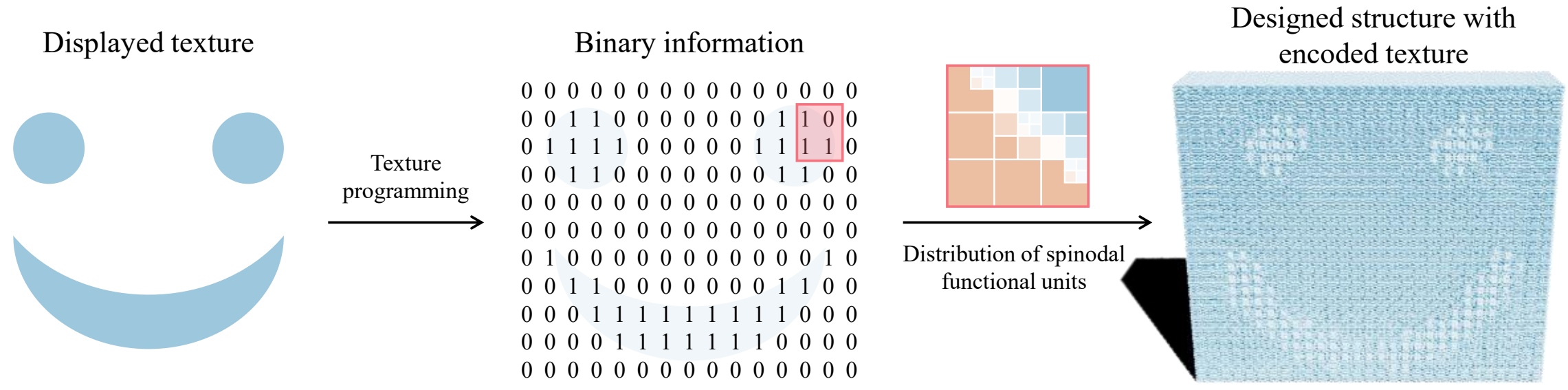

Figure S9. The design process of textured metamaterial with Emoji texture: By converting the texture into binary information, and using the proposed method to achieve smoother property transition for anti-aliasing.

### S2.4 Design of mechanical cloaking metamaterials

Mechanical cloaking metamaterials aim to rectify the deformation fields of the background region in order to make the internal void invisible to external observers. In this design process, gradient-based optimization is needed to determine the anisotropic stiffness matrix at the cloak region. Mechanical cloaking metamaterial aims to make internal voids or defects "invisible" from external static loads, by preserving similar deformation behavior as if no internal void exists. This concept is similar to optical cloaking, but applied to mechanical structures.

### S2.4.1 Design variable settings

Consider a cubic structure with a spherical void of radius $R_{\text{in}}$ at the center. The spherical region within a specified distance to the void is defined as the cloak region, with an outer radius of $R_{\text{out}}$, as shown in Figure S10. The structure of the cloak region is regarded as an anisotropic elastic body to compensate the nonlinear behavior induced by internal void. In the local principal axis coordinate system $(x', y', z')$, its local stiffness tensor in the Voigt notation is:

$$C_0 = \begin{bmatrix} C_{11} & C_{12} & C_{13} & C_{14} & C_{15} & C_{16} \\ C_{12} & C_{22} & C_{23} & C_{24} & C_{25} & C_{26} \\ C_{13} & C_{23} & C_{33} & C_{34} & C_{35} & C_{36} \\ C_{14} & C_{24} & C_{34} & C_{44} & C_{45} & C_{46} \\ C_{15} & C_{25} & C_{35} & C_{45} & C_{55} & C_{56} \\ C_{16} & C_{26} & C_{36} & C_{46} & C_{56} & C_{66} \end{bmatrix} \tag{S23}$$

The orientation of the local principal axis system relative to the global coordinate system $(x, y, z)$ can be described by three Euler angles $(\theta_1, \theta_2, \theta_3)$, and its rotation matrix $\boldsymbol{R}$ is:

$$\boldsymbol{R}_x(\theta_1) = \begin{bmatrix} 1 & 0 & 0 \\ 0 & \cos\theta_1 & -\sin\theta_1 \\ 0 & \sin\theta_1 & \cos\theta_1 \end{bmatrix} \tag{S24}$$

$$\boldsymbol{R}_y(\theta_2) = \begin{bmatrix} \cos\theta_2 & 0 & \sin\theta_2 \\ 0 & 1 & 0 \\ -\sin\theta_2 & 0 & \cos\theta_2 \end{bmatrix} \tag{S25}$$

$$\boldsymbol{R}_z(\theta_3) = \begin{bmatrix} \cos\theta_3 & -\sin\theta_3 & 0 \\ \sin\theta_3 & \cos\theta_3 & 0 \\ 0 & 0 & 1 \end{bmatrix} \tag{S26}$$

After rotation, the transformed stiffness tensor $C$ in the global coordinate system becomes:

$$C = \boldsymbol{T}(\boldsymbol{R}) C_0 \boldsymbol{T}^T(\boldsymbol{R}) \tag{S27}$$

where, $\boldsymbol{T}(\boldsymbol{R})$ is a $6 \times 6$ transformation matrix in Voigt notation with respect to the initial rotation matrix $\boldsymbol{R}$, which can be derived analytically based on the transformed constitutive relationship $\boldsymbol{T}(\boldsymbol{R})\sigma = \boldsymbol{T}(\boldsymbol{R})C_0\boldsymbol{T}^T(\boldsymbol{R})\varepsilon$:

$$\varepsilon'_{pq} = \boldsymbol{R}_{pm}\boldsymbol{R}_{qn}\varepsilon_{mn} \tag{S28}$$

Both sides are represented by Voigt vectors, resulting in a linear relationship:

$$\varepsilon' = \boldsymbol{T}(\boldsymbol{R})\varepsilon \tag{S29}$$

Therefore, the transformation matrix $\boldsymbol{T}$ in Voigt notation can be derived:

$$\boldsymbol{T}(\boldsymbol{R}) = \begin{bmatrix} r_{11}^2 & r_{12}^2 & r_{13}^2 & 2r_{12}r_{13} & 2r_{11}r_{13} & 2r_{11}r_{12} \\ r_{21}^2 & r_{22}^2 & r_{23}^2 & 2r_{22}r_{23} & 2r_{21}r_{23} & 2r_{21}r_{22} \\ r_{31}^2 & r_{32}^2 & r_{33}^2 & 2r_{32}r_{33} & 2r_{31}r_{33} & 2r_{31}r_{32} \\ r_{21}r_{31} & r_{22}r_{32} & r_{22}r_{33} & r_{22}r_{33} + r_{23}r_{32} & r_{21}r_{33} + r_{23}r_{31} & r_{21}r_{32} + r_{22}r_{31} \\ r_{11}r_{21} & r_{12}r_{32} & r_{13}r_{33} & r_{12}r_{32} + r_{13}r_{32} & r_{11}r_{23} + r_{13}r_{31} & r_{11}r_{32} + r_{12}r_{31} \\ r_{11}r_{22} & r_{12}r_{22} & r_{13}r_{23} & r_{12}r_{23} + r_{13}r_{22} & r_{11}r_{23} + r_{13}r_{21} & r_{11}r_{22} + r_{12}r_{21} \end{bmatrix} \tag{S30}$$

$$\boldsymbol{R} = \begin{bmatrix} r_{11} & r_{12} & r_{13} \\ r_{21} & r_{22} & r_{23} \\ r_{31} & r_{32} & r_{33} \end{bmatrix} \tag{S31}$$

The Voigt stiffness of the local principal axis is $C_0$, and the scaling vector of the principal axis using the Voigt component scale is:

$$\boldsymbol{s} = [s_1, s_2, s_3, s_2s_3, s_1s_3, s_1s_2]^T \tag{S32}$$

Let $C_{loacl} = \mathrm{diag}(\boldsymbol{s})C_0$, then the stiffness tensor can be defined as:

$$C = \boldsymbol{T}(\boldsymbol{R})C_{loacl}\boldsymbol{T}^T(\boldsymbol{R}) \tag{S33}$$

The displacement field of the reference structure (a complete cube composed of homogeneous units) under the same load is denoted as $\boldsymbol{u}_{ref}(\boldsymbol{x})$. The objective of the cloak optimization is to make the mechanical response of the designed structure consistent with the reference field, that is, to minimize the relative displacement error in the surrounding region $O$:

$$J = \frac{\int_O \|\boldsymbol{u}(\boldsymbol{x}) - \boldsymbol{u}_{ref}(\boldsymbol{x})\|^2 dV}{\int_O \|\boldsymbol{u}_{ref}(\boldsymbol{x})\|^2 dV} \tag{S34}$$

Subsequently, the linear elastic equation can be solved using finite element discretization:

$$K(C(\boldsymbol{p}))\boldsymbol{u} = f \tag{S35}$$

where $K$ represents the global stiffness matrix assembled by the element stiffness tensor $C(\boldsymbol{p})$, $f$ is the load vector, and $\boldsymbol{p}$ denotes all the design variables $\boldsymbol{p} = [\theta_1, \theta_2, \theta_3, s_1, s_2, s_3]$, $s_1, s_2, s_3$ represents the scaling size along three axes. The gradient of the objective function with respect to the design variables can be calculated through the adjoint method based on the following procedure:

(a) Solve the forward problem to obtain the displacement field $\boldsymbol{u}$;

(b) Construct the adjoint equation based on the objective function:

$$K^T \boldsymbol{\lambda} = \frac{\partial J}{\partial \boldsymbol{u}} \tag{S36}$$

Solve the adjoint variables $\boldsymbol{\lambda}$;

(c) Calculate the gradient:

$$\frac{dJ}{dp_k} = \frac{\partial R}{\partial \boldsymbol{p}_k} - \boldsymbol{\lambda}^T \left(\frac{\partial K}{\partial p_k}\right) \boldsymbol{u} \tag{S37}$$

Finally, the iterative optimization process is as follows:

**Step 1.** The initial stiffness $C_0$ in the design space is uniformly distributed, which is used to calculate the reference displacement field $\boldsymbol{u}_{\mathrm{ref}}$.

**Step 2.** Set the design variable $\boldsymbol{p}(\boldsymbol{x})$ within the cloaking region, and calculate the corresponding stiffness tensor $C(\boldsymbol{x})$ under corresponding rotation and scaling operation from

$\boldsymbol{p}(\boldsymbol{x})$.

**Step 3.** Obtain the corresponding displacement field $\boldsymbol{u}(\boldsymbol{x})$ under $\boldsymbol{p}(\boldsymbol{x})$.

**Step 4.** Calculate the objective function $J(\boldsymbol{p})$ based on the current results.

**Step 5.** Calculate the gradient through sensitivity analysis (the adjoint method): $\frac{\partial J}{\partial \boldsymbol{p}}$.

**Step 6.** Update the design variables using gradient-based constrained optimization methods.

**Step 7.** Repeat Steps 3-6 until the objective function converges or the maximum number of iterations is reached.

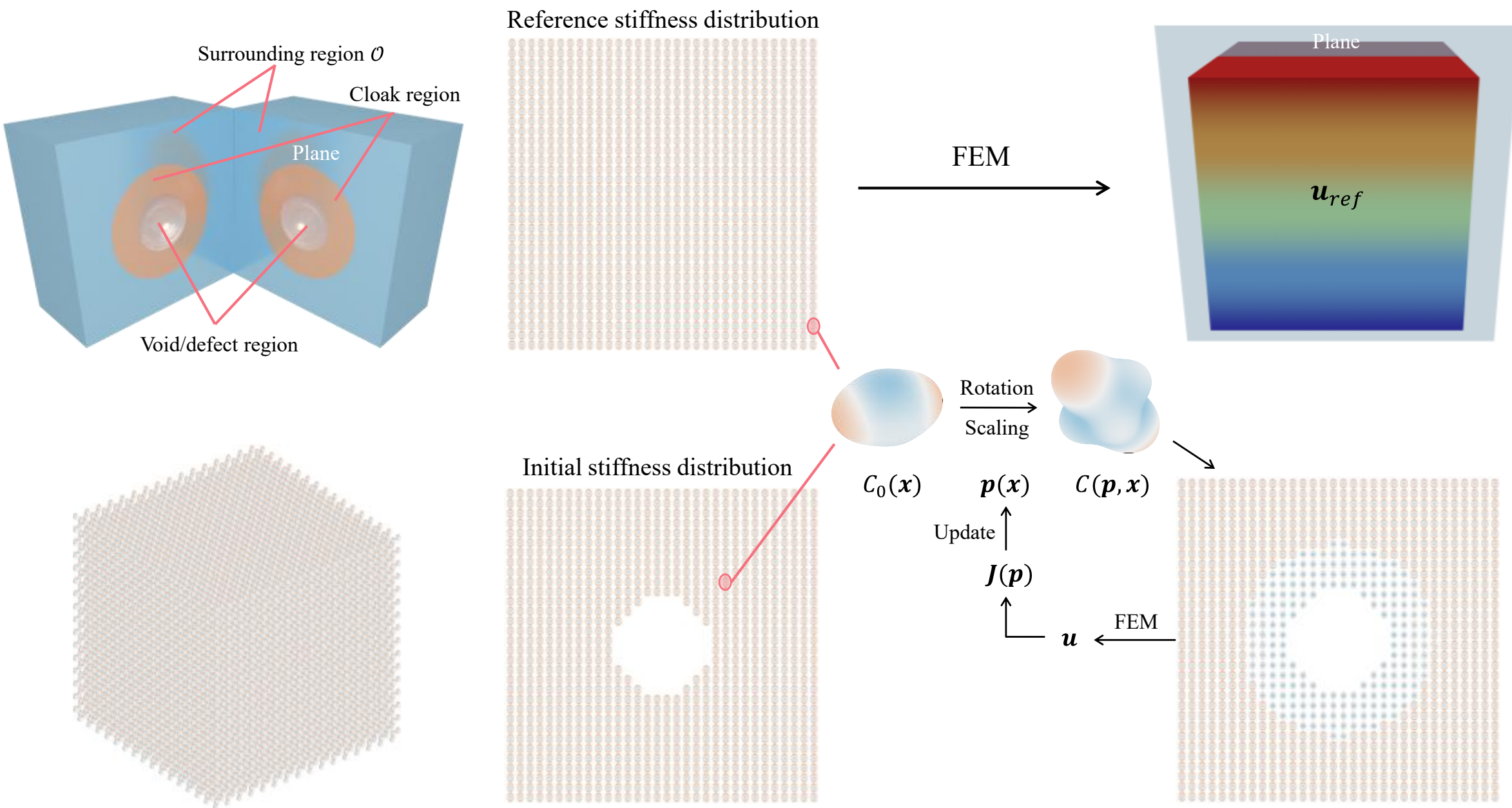

Figure S10. Schematic diagram of mechanical cloaking metamaterial design.

For detailed implementation, this case study adopted a 60*60*60mm cubic volume as design space, and specified a spherical void region with $R_{in} = 18\text{mm}$ and the cloak region with $R_{out} = 36\text{mm}$. A displacement of -7mm is applied to the top surface as the boundary condition. The initial stiffness tensor is uniformly set to $C_0$ with the corresponding parameters

as follows:

$$E_x = 5 \;\; E_y = 3 \;\; E_z = 2$$

$$v_{xy} = 0.3 \;\; v_{yz} = 0.3 \;\; v_{xz} = 0.3 \tag{S38}$$

$$G_{xy} = 1 \;\; G_{yz} = 1 \;\; G_{xz} = 1$$

Therefore, compliance tensor $S_0 = inv(C_0)$ (6×6) for initial input can be obtained as follows:

$$S_0 = \begin{bmatrix} \frac{1}{E_x} & -\frac{v_{yx}}{E_y} & -\frac{v_{xz}}{E_z} & 0 & 0 & 0 \\ -\frac{v_{xy}}{E_x} & \frac{1}{E_y} & -\frac{v_{yz}}{E_z} & 0 & 0 & 0 \\ -\frac{v_{xz}}{E_x} & -\frac{v_{yz}}{E_y} & \frac{1}{E_z} & 0 & 0 & 0 \\ 0 & 0 & 0 & \frac{1}{G_{yz}} & 0 & 0 \\ 0 & 0 & 0 & 0 & \frac{1}{G_{xz}} & 0 \\ 0 & 0 & 0 & 0 & 0 & \frac{1}{G_{xy}} \end{bmatrix} \tag{S39}$$

**S2.4.2 Sensitivity Analysis**

In the numerical optimization design process of a mechanical cloak, the design variables (such as rotation angle $\theta_i$ and the scaling factor $s_i$ to change the stiffness tensor $C(\boldsymbol{p})$) will affect the stress and displacement distribution of the entire structure. The optimization objective function $J$ depends on the displacement field $\boldsymbol{u}$, which is the solution of the PDE. Therefore, the problem becomes a PDE-constrained optimization problem:

$$\min_{\boldsymbol{p}} J\,(\boldsymbol{u}(\boldsymbol{p}), \boldsymbol{p}) \;\; s.t. \;\; \nabla \cdot (C(\boldsymbol{p}) : \nabla^s \boldsymbol{u}) = 0 \tag{S40}$$

To iteratively solve this problem, it is necessary to know the derivatives of the objective function, which can be divided into two parts with respect to the design variables $\frac{\partial J}{\partial \boldsymbol{p}}$:

$$\frac{dJ}{d\boldsymbol{p}} = \frac{\partial J}{\partial \boldsymbol{p}} + \frac{\partial J}{\partial \boldsymbol{u}}\frac{\partial \boldsymbol{u}}{\partial \boldsymbol{p}} \tag{S41}$$

The first term is the explicit term directly derived from the partial derivative of $J$ with respect to $\boldsymbol{p}$. The second term is the implicit term. Since $\boldsymbol{u}$ depends on $\boldsymbol{p}$, its derivatives must be derived through the PDE. Therefore, the key to solving the second term lies in the adjoint method.

The following part elaborates on the derivation of the adjoint sensitivity in elasticity mechanics. The elastic equilibrium equation can be written as:

$$\nabla \cdot (C(\boldsymbol{p}) : \nabla^s \boldsymbol{u}) = 0 \;\; in\, \Omega \tag{S42}$$

in which the boundary conditions are:

$$\begin{cases} \boldsymbol{u} = \boldsymbol{u}_0 \; on \; \Gamma_u \\ (C(\boldsymbol{p}) : \nabla^s \boldsymbol{u}) \cdot \boldsymbol{n} = \boldsymbol{t}_0 \; on \; \Gamma_t \end{cases} \tag{S43}$$

where, $C(\boldsymbol{p})$ is the stiffness tensor determined by the design variables, $\nabla^s \boldsymbol{u}$ is the symmetric gradient (strain tensor), which indicates the magnitude and direction of local deformation, $\Gamma_u$ is the displacement boundary condition. $\Gamma_t$ is the loading condition. $\boldsymbol{n}$ is the unit normal vector indicating the direction of the force acting on the boundary ($\boldsymbol{\sigma} \cdot \boldsymbol{n}$). $\boldsymbol{u}$ represents the actually measurable displacement distribution. Here, we optimize for rectifying the displacement field $\boldsymbol{u}$ in the surrounding region to align with the reference displacement field. The strain vector and stress vector can then be determined accordingly:

$$\varepsilon = \nabla^s \boldsymbol{u} \tag{S44}$$

$$\boldsymbol{\sigma} = C : \varepsilon \tag{S45}$$

This constitutive equation indicates that the stress and strain relationship is controlled by the stiffness tensor. In this case study, the cloaking effect is achieved by optimizing $C(\boldsymbol{p})$ with respect to the following equation:

$$\nabla \cdot \boldsymbol{\sigma} = 0 \tag{S46}$$

Let the displacement of the reference structure be $\boldsymbol{u}_{ref}$ and the displacement of the optimized structure be $\boldsymbol{u}$. The objective function at the surrounding region (denoted as $O$) is shown in Equation S25.

For sensitivity analysis using the adjoint method, we introduce the Lagrange multiplier $\boldsymbol{\lambda}$, and define the Lagrangian functional:

$$\mathcal{L}(\boldsymbol{u}, \boldsymbol{\lambda}, C) = J(\boldsymbol{u}) + \boldsymbol{\lambda}^T (f - K(C)\boldsymbol{u}) \tag{S47}$$

By substituting $J(\boldsymbol{u})$ into Equation S38, we can get:

$$\mathcal{L}(\boldsymbol{u}, \boldsymbol{\lambda}, C) = \frac{\int_O \left\| \boldsymbol{u}(x) - \boldsymbol{u}_{ref}(x) \right\|^2 dV}{\int_O \left\| \boldsymbol{u}_{ref}(x) \right\|^2 dV} + \boldsymbol{\lambda}^T (f - K(C(\boldsymbol{p}))\boldsymbol{u}) \tag{S48}$$

The derivative of $\mathcal{L}$ with respect to the displacement $\boldsymbol{u}$ is then set to zero:

$$\frac{\partial \mathcal{L}}{\partial \boldsymbol{u}} = 0 \;\; \Rightarrow \;\; K^T \boldsymbol{\lambda} = \frac{\partial J}{\partial \boldsymbol{u}} \tag{S49}$$

Next, for the convenience of writing, we define the objective function $J$ in the discrete form:

$$J(\boldsymbol{p}) = \frac{\boldsymbol{r}^T \boldsymbol{r}}{{\boldsymbol{u}_{ref}^O}^T \boldsymbol{u}_{ref}^O} \tag{S50}$$

$$\boldsymbol{r} = M_O(\boldsymbol{u} - \boldsymbol{u}_{ref}) \tag{S51}$$

where $M_O$ is a mask matrix that takes the value of 1 only at the surrounding region. Make the denominator of $J$ equal to $D = {\boldsymbol{u}_{ref}^{O}}^{T} \boldsymbol{u}_{ref}^{O}$, then:

$$\frac{\partial J}{\partial \boldsymbol{u}} = \frac{2}{D} M_O^T \boldsymbol{r} \tag{S52}$$

making the associated equation expressed as:

$$K^T \boldsymbol{\lambda} = \frac{2}{D} M_O^T \boldsymbol{r} \tag{S53}$$

when $K$ is symmetric, it is equivalent to:

$$K(\boldsymbol{p}) \boldsymbol{\lambda} = \frac{2}{D} M_O^T \boldsymbol{r} \tag{S54}$$

When both $\boldsymbol{u}$ and $\boldsymbol{\lambda}$ satisfy the state and adjoint equations ($\frac{\partial \mathcal{L}}{\partial \boldsymbol{\lambda}} = 0$), the derivative of the objective with respect to the design variable $\boldsymbol{p}$ is:

$$\frac{dJ}{d\boldsymbol{p}} = \frac{\partial \mathcal{L}}{\partial \boldsymbol{p}} = -\boldsymbol{\lambda}^T \frac{\partial K(C(\boldsymbol{p}))}{\partial \boldsymbol{p}} \boldsymbol{u} + \frac{\partial}{\partial \boldsymbol{p}} \left(\frac{\boldsymbol{r}^T \boldsymbol{r}}{D}\right) \tag{S55}$$

If there is no explicit relationship between $\boldsymbol{u}_{ref}$ and $\boldsymbol{p}$:

$$\frac{\partial}{\partial \boldsymbol{p}} \left(\frac{\boldsymbol{r}^T \boldsymbol{r}}{D}\right) = 0 \tag{S56}$$

Expanded by each element results in:

$$\frac{dJ}{d\boldsymbol{p}_e} = -{\boldsymbol{\lambda}_e}^T \frac{\partial K_e(\boldsymbol{p}_e)}{\partial \boldsymbol{p}_e} u_e \tag{S57}$$

From Equation S27, the unit stiffness $K_e$ is derived as:

$$K_e(\boldsymbol{p}_e) = \int_{\Omega_e} \boldsymbol{B}^T C_e(\boldsymbol{p}_e) \boldsymbol{B} \, d\Omega \tag{S58}$$

where, $\boldsymbol{B}$ is the strain-displacement matrix, representing the linear transformation from the node displacement to the element strain:

$$\varepsilon(\boldsymbol{u})_e = \boldsymbol{B} \, \boldsymbol{u}_e, \quad \varepsilon(\boldsymbol{\lambda})_e = \boldsymbol{B} \, \boldsymbol{\lambda}_e, \tag{S59}$$

Therefore,

$$\frac{\partial K_e(\boldsymbol{p}_e)}{\partial \boldsymbol{p}_e} = \int_{\Omega_e} \boldsymbol{B}^T \frac{\partial C_e(\boldsymbol{p}_e)}{\partial \boldsymbol{p}_e} \boldsymbol{B}\, d\Omega \quad \text{(S60)}$$

Finally,

$$\frac{dJ}{d\boldsymbol{p}_e} = -\int_{\Omega_e} (\boldsymbol{B}\boldsymbol{u}_e)^T \left(\frac{\partial C_e(\boldsymbol{p}_e)}{\partial \boldsymbol{p}_e}\right)(\boldsymbol{B}\boldsymbol{\lambda}_e)\, d\Omega \quad \text{(S61)}$$

The approximation of integral points for the conventional tetrahedron can be expressed as:

$$\frac{dJ}{d\boldsymbol{p}_e} \approx -V_e \varepsilon(\boldsymbol{u})_e^T \left(\frac{\partial C_e(\boldsymbol{p}_e)}{\partial \boldsymbol{p}_e}\right) \varepsilon(\boldsymbol{\lambda})_e \quad \text{(S62)}$$

As mentioned above, the design variables $\boldsymbol{p}$ consist of two sets of parameters, i.e. scaling factors and rotation angles. Therefore, the derivatives of the design variables $\boldsymbol{p}$ are divided into two categories: the derivatives with respect to scaling factor $s_j$ and the derivatives with respect to rotation angle $\theta_i$.

Because $C_e = \boldsymbol{T} C_{loacl} \boldsymbol{T}^T$ and $C_{loacl} = diag(\boldsymbol{s}) C_0$, we can get:

$$\frac{\partial C_e}{\partial s_j} = \boldsymbol{T} \frac{\partial C_{loacl}}{\partial s_j} \boldsymbol{T}^T \quad \text{(S63)}$$

$$\frac{\partial C_{loacl}}{\partial s_j} = diag(e_j) C_0 \quad \text{(S64)}$$

where $\mathrm{e_j}$ represents the position of the jth unit vector in the Voigt form, then:

$$\frac{dJ}{ds_j^{(e)}} \approx -V_e \varepsilon(\boldsymbol{u})_e^T \left(\boldsymbol{T} \frac{\partial C_{loacl}}{\partial s_j} \boldsymbol{T}^T\right) \varepsilon(\boldsymbol{\lambda})_e \quad \text{(S65)}$$

The derivative of the rotation angle $\theta_i$, according to the chain rule:

$$\frac{\partial C_e}{\partial \theta_i} = \frac{\partial \boldsymbol{T}}{\partial \theta_i} C_{loacl} \boldsymbol{T}^T + \boldsymbol{T} C_{loacl} \frac{\partial \boldsymbol{T}^T}{\partial \theta_i} \quad \text{(S66)}$$

where $\frac{\partial \boldsymbol{T}}{\partial \theta_i}$ can be obtained by taking the partial derivative of each $\boldsymbol{T}$ component with respect to the rotation element $r_{ij}$, and then multiplying by $\frac{\partial r_{ij}}{\partial \theta_i}$, which comes from $\frac{\partial \boldsymbol{R}}{\partial \theta_i}$, whose analytical expression is given by the selected Euler angle sequence. Therefore, $\frac{\partial \boldsymbol{T}}{\partial \theta_i}$ can be obtained as an analytical expression (without the need for numerical differentiation).

## S3 Manufacturing process

### S3.1 Additive manufacturing equipment and material

Most of the designed components in this work were manufactured using the Anycubic Photon D2 digital light processing (DLP) 3D printer for its high accuracy in fabricating stochastic microstructure. The polymer liquid, which is sensitive to ultraviolet light, was solidified layer by layer through a projector, thereby forming 3D printed objects. The 3D printing process is as follows:

**Step 1.** Prepare the photopolymer resin.

**Step 2.** The forming platform descends to the bottom of the resin tank, leaving a layer thickness (50 μm) of gap.

**Step 3.** The projector projects the current sliced image onto the resin layer: the area exposed to ultraviolet light instantaneously polymerizes and solidifies to form a solid layer.

**Step 4.** Repeat the above steps until all layers are completed.

The build volume of the Anycubic Photon D2 3D printer is $130.56 \times 73.44 \times 165$ mm, with a layer internal resolution of $51\mu\text{m}$ (i.e., the pixel size is $51 \times 51\mu\text{m}$). The layer thickness used in this work is $50\mu\text{m}$. All models were fabricated using ANYCUBIC Tough Resin Ultra. The exposure time for each layer was 2.3 seconds, the lighting-off time is 1 seconds, the Z-axis lifting height is 5 mm, the lifting speed is 2 mm/s, and the retreating speed is 3mm/s. The properties of the materials used are shown in Table S1.

Table S1 Specifications of the applied resin.

| Properties | Value |
| --- | --- |
| Density | 1.05~1.25 $g/cm^3$ |
| Bending strength | 15~25 MPa |
| Elongation at break | 70%~80% |
| Tensile strength | 20~28 MPa |
| Bending modulus | 350~800 MPa |

### S3.2 Post-processing

In order to fabricate physical models with expected design property, this study sets up the complete post-processing procedure as shown in the Figure S11, aiming to minimize the impact of additive manufacturing on the structural performance. The complete post-processing procedure is as follows:

**Step 1.** After printing, remove the printed piece from the platform.

**Step 2.** Use wash & curing machine with 95% medicinal alcohol for a 5-minute cleaning, and apply ultrasonic cleaning for 2 minutes.

**Step 3.** Use a commercial salad spinner to remove the residual resin/alcohol from the structure by centrifugal drying.

**Step 4.** Place the printed piece in the ANYCUBIC wash & cure curing machine for post curing with 3 minutes.

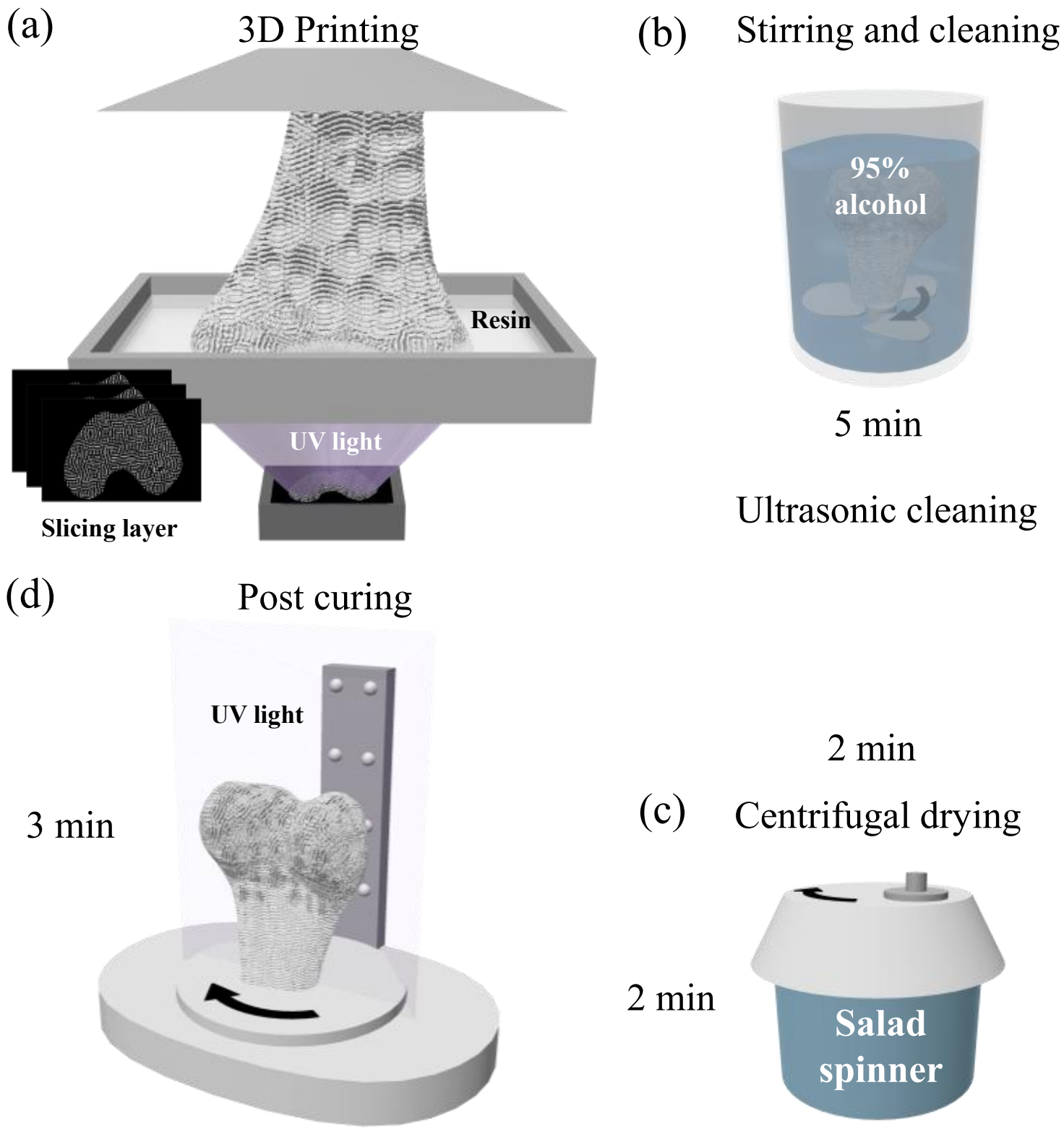

Figure S11. Fabrication procedures for 3D printed testing pieces: (a) 3D printing using DLP printer; (b) wash the residual resin on the samples with 95% medicinal alcohol, followed by a secondary ultrasonic cleaning; (c) centrifugal drying to eliminate the residual liquid resin/alcohol on the structure; (d) post-curing treatment.

## S4 Experimental setup

In this study, in order to verify the effectiveness of the proposed method in various applications, we conducted mechanical tests for the fabricated components using an INSTRON universal testing machine 5986.

For the test piece of the femur, the structure is in the size of 120mm*40mm*70mm with complex geometric profile. To ensure adequate contact between the structure and the testing machine, a preload force of 20N was set, compression was carried out with preloading rate at 1mm/min, and compression rate at 2mm/min until the total compression in Z direction reached 27mm.

For the testing piece of the mechanical cloak, we conducted similar mechanical test on the universal testing machine. To ensure adequate contact between the structure and the testing machine, a preload force of 2N was set, compression was carried out with preloading rate at 1mm/min, and compression rate at 1mm/min. Additionally, to facilitate comparison of the displacement fields, we utilized 3D digital image correlation technology (DIC). During the deformation process, the stereoscopic images were captured to estimate the global displacement of the sample surface. The experimental setup included an INSTRON universal testing machine and a HAYTHAM DV 2600-15 DIC system for displacement detection, as shown in Figure S12.

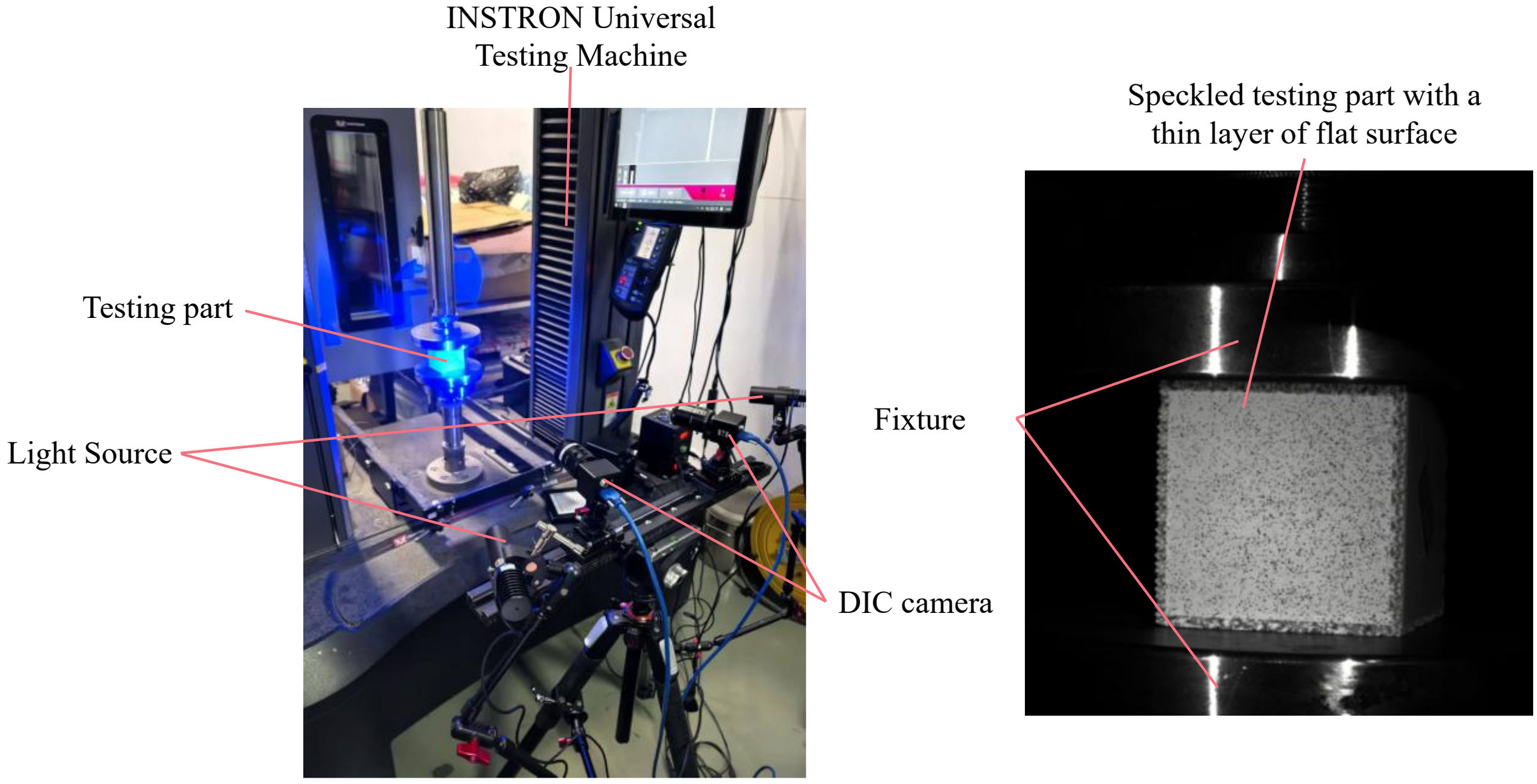

Figure S12. The experimental setup for testing mechanical cloaking metamaterials.

DIC calculates the displacement vector of each sub-region (subset) in the image by comparing the gray-scale distribution of the speckle pattern during the loading process, which serves as the "fingerprint" tracked by the algorithm. The texture features directly affect the accuracy of sub-pixel interpolation, and the spatial continuity of the displacement/stress field.

To enable detectable adhesion of speckle pattern onto the surface of the designed stochastic metamaterials, a thin planar layer (0.3mm thickness) was designed and fabricated to form a flat surface, allowing the speckles to be perfectly adhered on the structure for accurate evaluation.

## Video S1 The relationship between spherical harmonic coefficients and spherical functions

## Video S2 The relationship between Gaussian random field coefficients and spinodal structure

## Video S3 Textured metamaterials compression experiment

## Video S4 Mechanical cloaking compression experiment

The supplementary video has been uploaded to https://github.com/SiC35/Supplementary-video-of-Disordered-Continuity.